\documentclass[12pt]{iopart}
\usepackage{iopams}
\usepackage{subfigure}
\usepackage{graphicx}
\usepackage{color}
\usepackage{time}
\usepackage{enumerate}
\newcommand{\Schrodinger}{Schr{\"o}dinger}     
\newcommand{\ansatz}{ansatz}               
\newcommand{\dd}[1]{\,\mathrm{d}#1\,}
\newcommand{\ddd}[1]{\,\mathrm{d}^{2}#1\,}
\newcommand{\tint}{\!\int\!}                   
\newcommand{\eqref}[1]{(\ref{#1})}

\newcommand{\pdv}[2]{\frac{\partial #1}{\partial #2}}
\newcommand{\pddv}[2]{\frac{\partial^2 #1}{\partial #2^2}}
\newcommand{\qc}{\>,\quad}

\newcommand{\notag}{\nonumber}                 
\eqnobysec                                     
\usepackage{hyperref}          
\hypersetup{
    pdfnewwindow=true,         
    colorlinks=true,           
    linkcolor=red,             
    citecolor=blue,            
    filecolor=green,           
    urlcolor=cyan              
}
\begin{document}
\hfill LA-UR-22-25694

\hfill \today, \now \ PST

\vspace{10pt}
\title[2D-NLSE stability]{Stability of exact solutions of the $(2+1)$-dimensional nonlinear \Schrodinger\ equation with arbitrary nonlinearity parameter $\kappa$}
%
%
\author{Fred Cooper$^{1,2}$, Avinash Khare$^{3}$, Efstathios G. Charalampidis$^{4}$, John F. Dawson$^{5}$, and Avadh Saxena$^{2}$}
%
%
\address{$^1$The Santa Fe Institute, 
   1399 Hyde Park Road, 
   Santa Fe, NM 87501,
   United States of America}
\address{$^2$Center for Nonlinear Studies and Theoretical Division,
   Los Alamos National Laboratory,
   Los Alamos, NM 87545,
   United States of America}
\address{$^3$Physics Department, 
   Savitribai Phule Pune University, 
   Pune 411007, India}
\address{$^{4}$Mathematics Department, 
   California Polytechnic State University, 
   San Luis Obispo, CA 93407-0403, 
   United States of America}   
\address{$^5$Department of Physics,
   University of New Hampshire,
   Durham, NH 03824,
   United States of America}
   %
%
\ead{cooper@santafe.edu, khare@physics.unipune.ac.in, echarala@calpoly.edu, john.dawson@unh.edu, avadh@lanl.gov}
%
%
%
\begin{abstract}
In this work, we consider the nonlinear Schr\"odinger equation (NLSE) in $2+1$ dimensions with arbitrary nonlinearity exponent $\kappa$ in the presence of an external confining potential. Exact solutions to the system are constructed, and their stability over their ``mass'' (i.e., the $L^2$ norm) and the parameter $\kappa$ is explored. We observe both theoretically and numerically that the presence of the confining potential leads to wider domains of stability over the parameter space compared to the unconfined case. Our analysis suggests the existence of a stable regime of solutions for all $\kappa$ as long as their mass is less than a critical value $M^{\ast}(\kappa)$. Furthermore, we find that there are two different critical masses, one corresponding to width perturbations and the other one to translational perturbations. The results of Derrick's theorem are also obtained by studying the small amplitude regime of a four-parameter collective coordinate (4CC) approximation. A numerical stability analysis of the NLSE shows that the instability curve $M^{\ast}(\kappa)$ vs. $\kappa$ lies below the two curves found by Derrick's theorem and the 4CC approximation. In the absence of the external potential, $\kappa=1$ demarcates the separation between the blowup regime and the stable regime. In this 4CC approximation, for $\kappa<1$, when the mass is above the critical mass for the translational instability, quite complicated motions of the collective coordinates are possible. Energy conservation prevents the blowup of the solution as well as confines the center of the solution to a finite spatial domain. We call this regime the ``frustrated'' blowup regime and give some illustrations. In an appendix, we show how to extend these results to arbitrary initial ground state solution data and arbitrary spatial dimension $d$. 
\end{abstract}
%
%
%
%
\maketitle
%
%
\section{\label{s:intro} Introduction}

The nonlinear \Schrodinger\ equation (NLSE) is an important model of mathematical physics, 
having applications in plasma physics~\cite{KonoSkoric}, nonlinear optics~\cite{KivsharAgrawal}, 
water waves~\cite{Dauxois2006,Ablowitz2011} and Bose-Einstein condensate physics~\cite{PitaevskiiStringari2003,PethickSmith2002}. 
The phenomenon of solitary wave  blowup~\cite{SulemSulem1999} for Gaussian initial conditions of 
the NLSE  as a function of $\kappa d$ ($\kappa$ is the nonlinearity exponent and $d$ is the number 
of spatial dimensions) has been studied in the past both numerically \cite{Rose-1988-207}  and 
in a time-dependent  Hartree approximation  \cite{Cooper-1992-184} with the result that for 
$\kappa d > 2$ initial Gaussian conditions lead to blowup and at $\kappa d =2$ there is a critical 
mass for this blowup of initial data to occur. The fact that there can be finite-time blowup in 
nonlinear problems such as the NLSE has been known for a long time using norm inequalities~\cite{Ball-1977}. 

Recently it has been shown~\cite{Antar-2013}  that if we assume some initial data for the NLSE, 
one can rig up an external potential so that the initial data is the $t=0$ value of an exact solution. 
These authors utilized the homotopy analysis method~\cite{Liao-1992,He-1999}  to generate the exact 
solution.  However, in retrospect, it is clear that  one can easily find the external potential that 
makes an initial condition an exact solution at all times, by assuming that the time dependence of 
the exact solution is given by $\rme^{-\rmi \omega t}.$ This method, which we will use here, can be 
generalized to arbitrary initial conditions and arbitrary dimension.  The fact that this initial
condition is now an exact solution allows us to study stability using various exact and approximate 
methodologies.  We can then directly determine how this particular confining potential changes  
the criterion for blowup of Gaussian initial data. 

For the NLSE without a confining potential,  whether initial Gaussian data on the wavefunction 
$\psi(x,t)$ leads to  blowup or collapse \cite{Sulem-2013} was controlled by whether $\kappa d$ 
is greater or less than two. At the special case $\kappa d =2$, blowup only occurs when the conserved 
$L^2$-norm of the initial pulse  $M = \int \rmd^dx \, |\psi|^{2}$ is greater than a critical value. 
When we add the particular confining potential that makes the Gaussian wavefunction an exact solution, 
we find that the response of the wavefunction to small perturbations is quite different. Confining 
ourselves in this paper to $d=2$, we find that although the $\kappa=1$ threshold value separates two 
regions, i.e., one where blowup is possible and one where it is not, the stability is now also 
controlled by two critical masses denoted hereafter as $M_{\mathrm{w}}$ and $M_{\mathrm{t}}$, and related 
to the onset of width and translational instabilities, respectively, of the wavefunction. 

Indeed, for $\kappa < 1+\sqrt{2}$, the translational instability occurs before the width instability. 
We find that for $\kappa <1$, the critical value for blowup to occur, there are several regions. When 
$M< M_{\mathrm{t}}, M_{\mathrm{w}}$, the solutions are linearly stable, and one is in the small oscillation 
regime for the width and for the position when we perturb the width and position slightly. However, 
when $M> M_{\mathrm{t}}, M_{\mathrm{w}}$ we are now in a new regime of \textit{frustrated blowup} as a 
result of energy conservation. In a 4-collective coordinate (4CC) approximation, the perturbed solution 
starts blowing up but then it gets frustrated at a critical time and very complicated behaviors of the 
collective coordinates (CCs) are possible. For $\kappa  > 1$ and $M < M_{\mathrm{w}}, M_{\mathrm{t}}$, 
we again have small oscillations when we perturb the initial conditions. The traditional type of blowup 
occurs when $M >  M_{\mathrm{w}}$~\cite{Derrick-1964}, and we show this in the 4CC variational approximation. 
We plot the energy landscape for both width and translational stability using a generalization of Derrick's 
theorem \cite{Dawson-2017}. The region of stability obtained from this analysis agrees with the small oscillation 
regime found in a 4CC approximation. This agreement between these two approaches was also found in a previous 
study of the $(1+1)$-dimensional NLSE in a P{\"o}schl-Teller external potential~\cite{Dawson-2017}.

The structure of the present paper is as follows. In Section~\ref{s:2NLSE},
we present our model together with the exact solution and the external potential 
we consider. We discuss about the associated Lagrangian dynamics and conserved
quantities in Sec.~\ref{s:2Ddynamics} while Sec.~\ref{s:Derrick} offers a systematic
study of the stability of the exact solution under width and translational perturbations
in view of Derrick's theorem. In Secs.~\ref{s:CCmethod} and~\ref{s:4CCevolutions},
we focus on a 4CC ansatz and present typical evolutions involving it therein. Section~\ref{s:stability}
discusses the spectral properties of the exact solutions to the NLSE in the realm of 
Bogoliubov-de Gennes (BdG) analysis. Finally, Sec.~\ref{s:conclusions} presents our 
conclusions.

%
%
\section{\label{s:2NLSE}The Model and Main Setup}

The~$(2+1)$-dimensional (one temporal and two spatial dimensions), 
nonlinear \Schrodinger\ equation (NLSE) in an external potential 
is given by:
\begin{equation}\label{NLSE-2D}
   \rmi \frac{\partial{\psi(\mathbf{r},t)}}{\partial t}
   +
   \nabla^{2} \psi(\mathbf{r},t)
   +
   g \, |\psi(\mathbf{r},t)|^{2 \kappa} \, \psi(\mathbf{r},t)
   =
   V(\mathbf{r}) \, \psi(\mathbf{r},t)\>,
\end{equation}
where $\psi(\mathbf{r},t)$ is a complex-valued wavefunction
(with $\mathbf{r} = (x,y)$ and $r=|\mathbf{r}|$), $g$ and $\kappa$ correspond to 
the nonlinearity strength and nonlinearity exponent, respectively,
and $V(\mathbf{r})$ is the external potential. If $V(\mathbf{r})\equiv 0$,
blowup of initial Gaussian data for $g > 0$ was studied at arbitrary 
$d$ both numerically and approximately in a time dependent Hartree 
approximation~\cite{Cooper-1992-184}. Here we would like to focus on 
the study of the stability of a Gaussian wavefunction when the latter 
is the {\em exact} solution of the NLSE [cf. Eq.~\eqref{NLSE-2D}] in a 
confining potential. To do this we will make use of recent work of Antar 
and Pamuk~\cite{Antar-2013}. 

\begin{figure*}[t]
   \centering
   \includegraphics[width=0.55\linewidth]{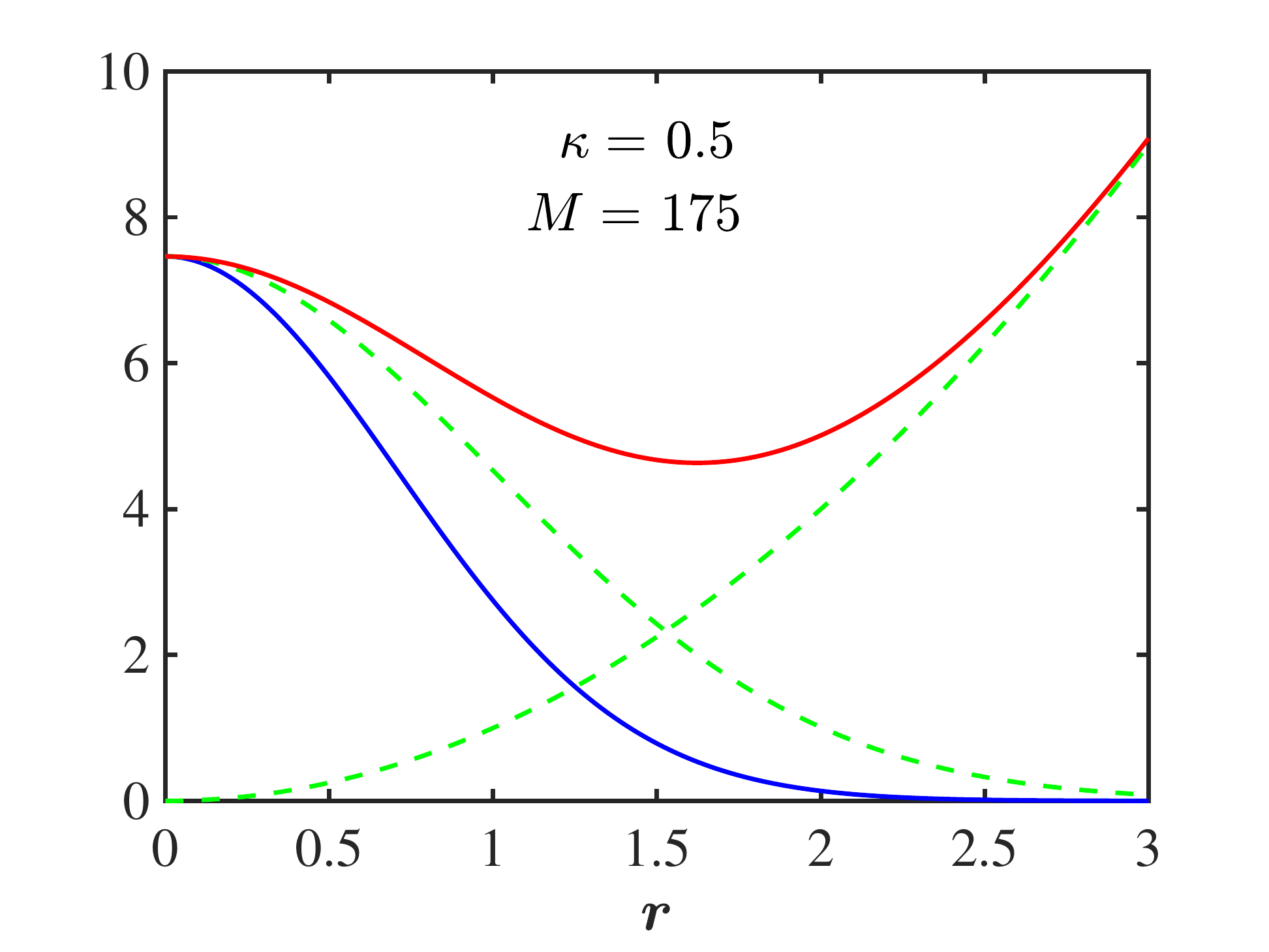}
   \caption{\label{f:Vrho}The density $\rho(r)$ (in blue) and confining 
   potential $V(r)$ (in red) as a function of $r$ for the case when $\kappa = 1/2$
   and $M = 175$ (see Table~\ref{t:plotdata}).  The dotted green lines 
   are plots of $V_1(r)$ and $V_2(r)$ for this case.}
\end{figure*}

Their results can be interpreted as a way of finding an external potential 
for the NLSE  which transforms the initial data for the NLSE into an exact 
solution of the  problem of the NLSE in an external potential by adding a 
particular time-dependent phase. Here we concern ourselves with the particular 
case of Gaussian initial data in order to compare with previous results in the 
absence of a confining potential
We thus start with the following ansatz:
\begin{equation}\label{assume1}
   \psi_0(\mathbf{r},t)
   =
   A_0 \, \rme^{ - r^2/(2 G_0) - \rmi \, \omega t},
   \quad
   A_0 > 0 \>,
\end{equation}
where $\omega$ stands for the phase, and we demand that Eq.~\eqref{assume1} is 
a solution to the NLSE in an external potential. Upon inserting Eq.~\eqref{assume1} 
into the left-hand-side (lhs) of Eq.~\eqref{NLSE-2D}, we find that the appropriate 
potential to make Eq.~\eqref{assume1} an exact solution is
\begin{eqnarray}
   V(\mathbf{r}) 
   &=&
   V_1(\mathbf{r}) + V_2(\mathbf{r})
   \qc
   \omega = 2 / G_0 \>,
   \notag \\
   V_1(\mathbf{r})
   &=&
   g \, A^{2 \kappa}_0 \, \rme^{- \kappa r^2/G_0 } \>,
   \notag \\
   V_2(\mathbf{r})
   &=&
   r^2 / G^2_0\>.
   \label{Vassume1}
\end{eqnarray}
%
A plot 
of the density $\rho(r) = |\psi_0(\mathbf{r},t)^2|$ and the potential 
$V(\mathbf{r})$ for the case when $\kappa = 1/2$ and $M = 175$ is shown 
in Fig.~\ref{f:Vrho}. The potential is a two-dimensional harmonic oscillator 
potential plus a Gaussian confining potential that is easy to construct experimentally 
using lasers. In the~\ref{s:extend}, we discuss how to determine the potential for 
arbitrary spherically symmetric (ground state) wavefunctions for arbitrary $d$,
and for arbitrary nonlinearity $\kappa$. 

%
%
%
%

%
%
\section{\label{s:2Ddynamics}Lagrangian dynamics in two spatial dimensions}

The Dirac action~\cite{Dirac-1930,Dirac-1934} that upon variation leads 
to the NLSE of Eq.~\eqref{NLSE-2D} for \emph{any} potential $V(x,y)$ 
is given by
\begin{eqnarray}
   \Gamma[\psi,\psi^{\ast}]
   &=&  
   \tint \dd{t} L[\psi,\psi^{\ast}]
   =
   \tint \dd{t} \{\, T[\psi,\psi^{\ast}] - H[\psi,\psi^{\ast}] \, \}\>,
   \label{GammaDef} 
   \\
   T[\psi,\psi^{\ast}]
   &=&
   \int_{-\infty}^{+\infty} \hspace{-1em} \ddd{x}  
      \Bigl [\,
         \frac{\rmi}{2}  \bigl( \psi^\ast (\partial_t \psi) 
         - 
         (\partial_t \psi^\ast) \psi \, \bigr)  
      \Bigr] \>,
   \label{Tdef} 
   \\
   H[\psi,\psi^{\ast}] 
   &=&
   \int_{-\infty}^{+\infty} \hspace{-0.75em} \ddd{x} 
   \Bigl [\,
     \, | \nabla \psi |^2
      -
      \frac{g}{\kappa+1} \, (\psi^\ast \psi)^{\kappa+1}
      +
      V \, | \psi |^2 \, 
   \Bigr ] \>.
   \label{Hdef}
\end{eqnarray} 
Here $\ddd{x} = \dd{x} \dd{y}$. For spherically symmetric wavefunctions,
the kinetic part of $H$ can be written in spherical coordinates as
\begin{equation}
   K[\psi,\psi^{\ast}]
   =
   2 \pi \int_0^\infty \hspace{-0.75em} r \dd{r}  
      \Big | \pdv{\psi}{r} \Big |^2
   =
   \int_{-\infty}^\infty \hspace{-1em} \dd{x} 
   \int_{-\infty}^\infty \hspace{-1em} \dd{y} 
   \Bigl [\, 
      \Big | \pdv{\psi}{x} \Big |^2
      +
      \Big | \pdv{\psi}{y} \Big |^2 \,
   \Bigr ] \>.
   \label{KII}
\end{equation}

%
%
\subsection{\label{ss:conserved} Conserved quantities}

From the equation of motion [cf.~Eq.~\eqref{NLSE-2D}], one finds that 
the $L^2$ norm of the wavefunction, called the mass $M$ hereafter, 
is conserved:
\begin{equation}\label{mass}
   M
   =
   \tint \ddd{x} |\psi(x,y,t)|^2,
\end{equation}
and for the exact solution of Eq.~\eqref{assume1}, the conserved mass
reduces into
\begin{equation}\label{e:massExact}
   M 
   = 
   2 \pi A_0^2 \int_{0}^{\infty} \!\!\! r \, \rme^{- r^2/G_0} \dd{r}
   =
   \pi \, G_0 \, A_0^2
   \qc
   A_0 = \sqrt{\frac{M}{\pi G_0}} \>.
\end{equation}
While studying the stability of the pertinent Gaussian waveforms, we will keep
the mass of the initial condition unchanged (over time $t$), although its initial 
width will be of the form of $G(0) = G_0/\beta = G_0 + \delta G_0$ (here, we adopt
the notation $G(0)\doteq G(t=0)$). 
The initial height of the Gaussian for the perturbed solution is then given by:
\begin{equation}
   A(0) = \sqrt{\frac{M}{\pi G(0) }} \>.
\end{equation}

The (total) energy given by Eq.~\eqref{Hdef} is also conserved, and for the exact solution, 
it is explicitly given by:
\begin{equation}
   \frac{E}{M} 
   = 
   \frac{2}{G_0}
   +
   \frac{g \, \kappa}{(\kappa + 1)^2} \, \Bigl ( \frac{M}{\pi \, G_0} \Bigr )^{\kappa} \>.
   \label{e:EnergyExact}
\end{equation}

%
%
\section{\label{s:Derrick}Derrick's theorem}

%
%
\subsection{\label{ss:Derrick2D}Width stability}

\begin{figure*}[t]
   \centering
   \subfigure[$M_{\mathrm{w}}$ (red), $M_{\mathrm{t}}$ (blue), and 
   numerical BdG analysis (black).]
   {\label{fig:Mcrit-a}
   \includegraphics[width=0.47\linewidth]{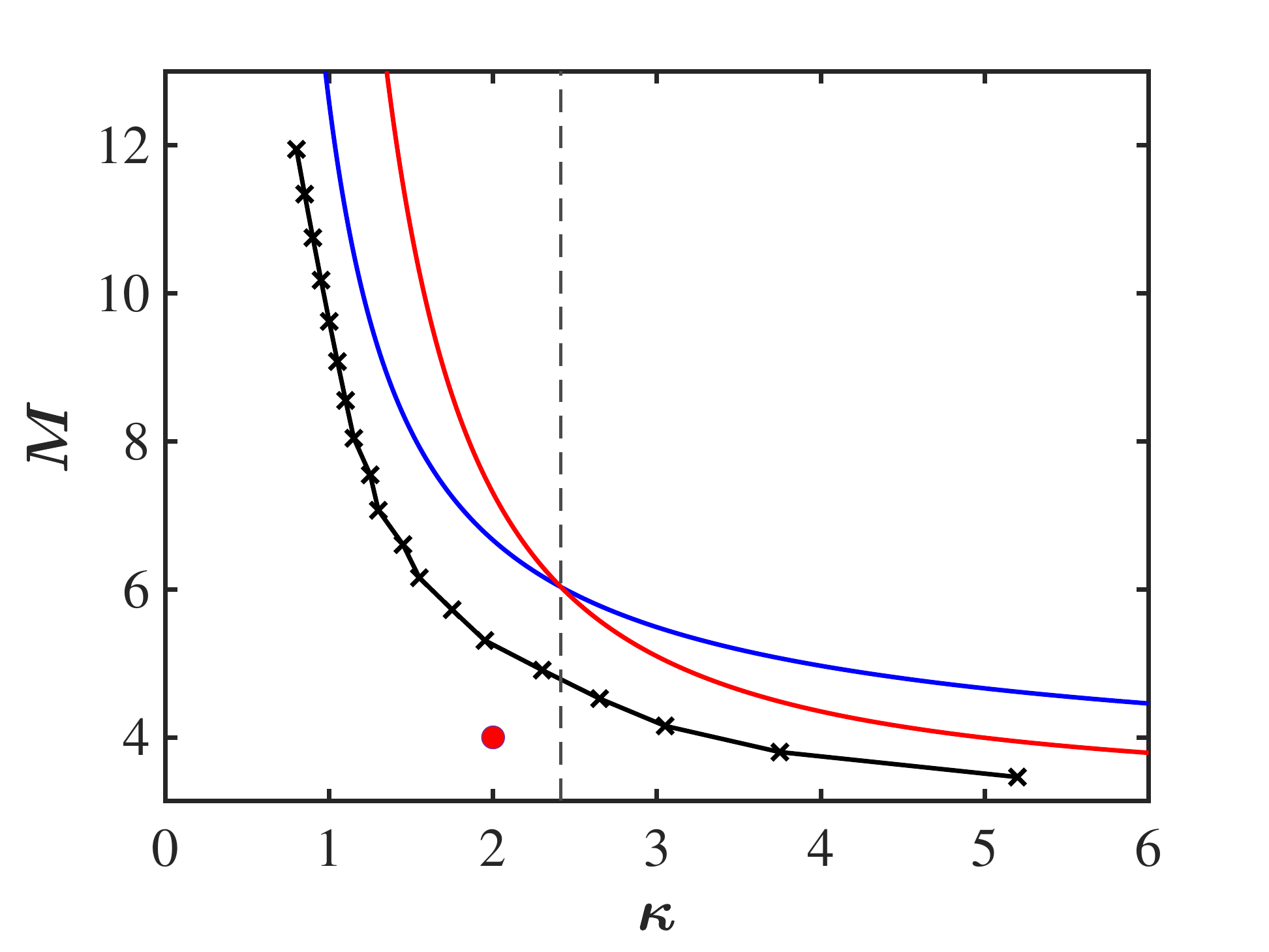} }
   \hspace{0.1em}
   \subfigure[Data points for 4CC simulations (see table \ref{t:plotdata}).]
   {\label{fig:Mcrit-b}
   \includegraphics[width=0.47\linewidth]{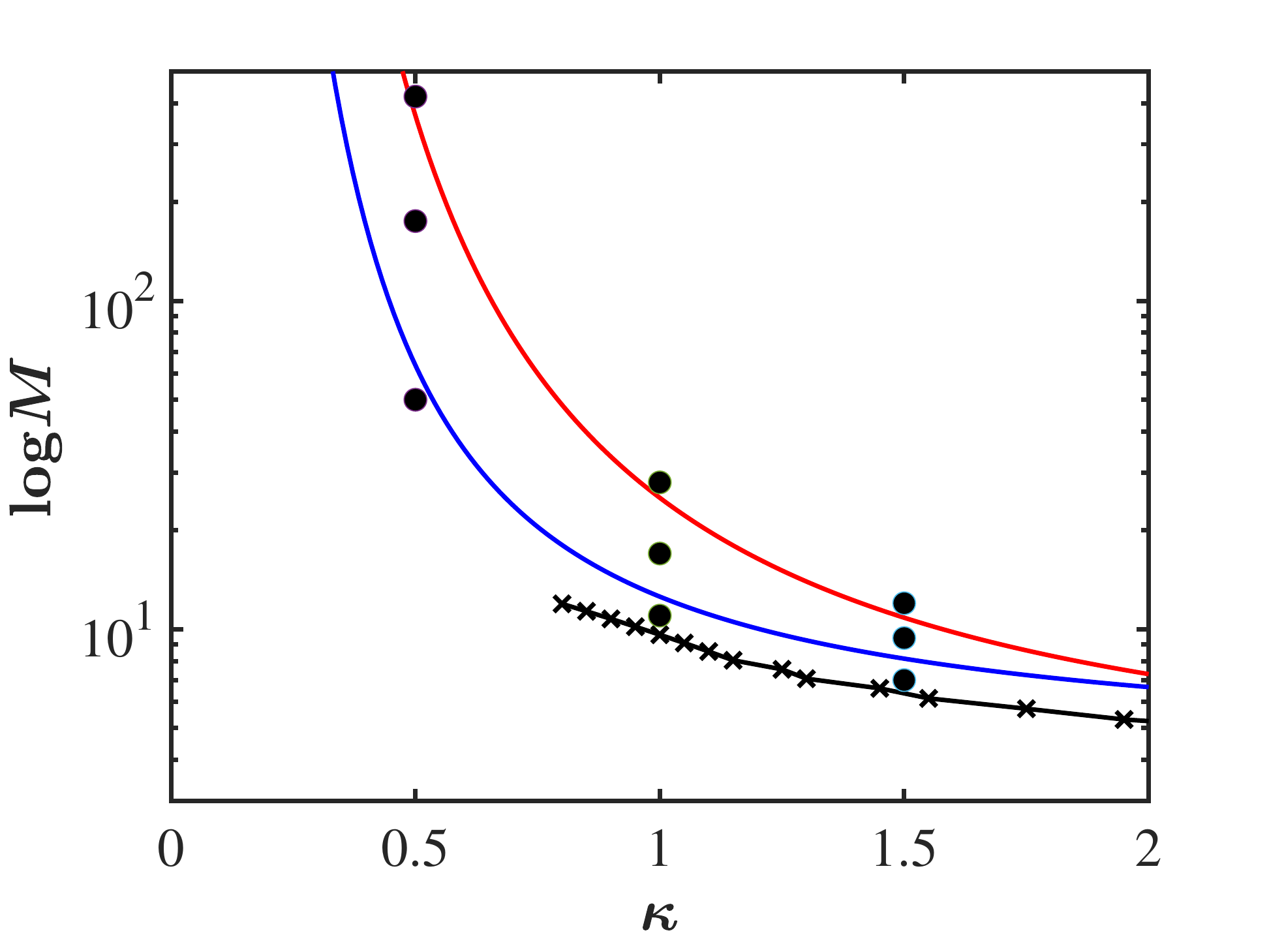} }
   \caption{\label{fig:Mcrit}The critical width mass $M_{\mathrm{w}}$ (red),
   critical translational mass $M_{\mathrm{t}}$ (blue), and BdG analysis (black)
   as a function of $\kappa$ (see also, Fig.~\ref{fig:Mstar3} in Section~\ref{s:stability}). 
   The red data point in panel (a) corresponds to the simulation
   shown in Fig.~\ref{f:Kappa-20}. The dashed vertical line is at the 
   intersection point $\kappa=1+\sqrt{2}$.}
\end{figure*}

First, we would like to see if the exact solution is stable to changes 
in the width while keeping the mass fixed. This is the criterion for stability 
due to Derrick~\cite{Derrick-1964}. It should be noted in passing that for 
$d=2$ and in the absence of the external potential, the solutions are unstable 
to changes in the width when $\kappa>1$. To that end, we set $r^2 \rightarrow \beta \, r^2$
(with $\beta$ being the rescaling parameter), and take the stretched wavefunction 
as
\begin{equation}\label{solwave}
   \tilde{\psi}(r,t)
   =
   \tilde{A} \, \rme^{- \beta \, r^2/(2 G_0) - \rmi \, \varphi(t)} \>,
\end{equation}
and examine what this transformation does to the Hamiltonian \eqref{Hdef}. 
Keeping the mass fixed, we arrive at 
\begin{equation}
   \tilde A^2 
   =  
   \frac{\beta M}{\pi G_0} 
   = 
   \beta \, A_0^2 \>,
\end{equation}
%
and thus, the density for the streched solution is given by:
\begin{equation}
   \tilde{\rho}(r) 
   = 
   |\tilde{\psi}(r,t)|^2 
   =
   \frac{\beta M}{\pi G_0} \rme^{- \beta \, r^2 / G_0}\>.
\end{equation}
To compare with previous work on blowup in the NLSE \cite{Dawson-2017}, 
we will eventually set $G_0=g=1$. We have that this solution contributes 
to the various components of the energy as follows:
\begin{eqnarray}
   H_1
   &=&
   2 \pi \int_0^\infty \hspace{-0.5em} r \dd{r} \Bigl |\pdv{\psi}{r} \Bigr |^2  
   = 
   \frac{M \beta}{G_0} \>,
   \label{e:H1-Derrick} 
   \\
   H_2 
   &=&      
   -
   \frac{g}{\kappa+1} \, 2 \pi \int_0^\infty \hspace{-0.5em} r \dd{r} 
   \tilde{\rho}^{\kappa+1}(r)
   = 
   - 
   \frac{g \,M}{(\kappa + 1)^2} \, 
   \Bigl [ \frac{\beta M}{\pi \, G_0} \Bigr ]^{\kappa} \>,
   \label{e:H2-Derrick}
   \\
   H_3 
   &=&
   2 \pi \int_0^\infty \hspace{-0.5em} r \dd{r}
   \tilde{\rho}(r) \, V_1(r)
   \notag \\
   &=&   
   g \, \beta \Bigl ( \frac{M }{\pi G_0} \Bigr )^{\kappa+1} \,
   2 \pi \int_0^\infty \hspace{-0.5em} r \dd{r}
   e^{-(\kappa+ \beta) \, r^2 / G_0} 
   =
   \frac{g \, \beta \, M }{\beta +\kappa } \,
   \Bigl [ \frac{M}{\pi \, G_0} \Bigr ]^{\kappa} \>,
   \label{e:H3-Derrick}
   \\
   H_4 
   &=&
   2 \pi \int_0^\infty \hspace{-0.5em} r \dd{r}
   \tilde{\rho}(r) \, V_2(r)
   \notag \\
   &=&
   \frac{\beta M }{\pi G_0} \, \frac{2 \pi}{G_0^2}  \int_0^\infty \hspace{-0.5em} r^3 \dd{r}  
   \rme^{- \beta \, r^2 / G_0} 
   = 
   \frac{M}{\beta \, G_0} \>.
   \label{e:H4-Derrick}
\end{eqnarray}
The Hamiltonian denoted by $H_{\mathrm{w}}$ in this case, is then
given by
\begin{equation}
   \frac{H_{\mathrm{w}}(\beta)}{M}  
   =
   \frac{1}{G_0} \, \Bigl [\, \beta + \frac{1}{\beta} \, \Bigr ]
   - 
   \frac{g}{(\kappa + 1)^2} \, 
   \Bigl [ \frac{\beta M}{\pi \, G_0} \Bigr ]^{\kappa}
   +
   \frac{g \, \beta}{\beta +\kappa } \,
   \Bigl [ \frac{M}{\pi \, G_0} \Bigr ]^{\kappa} \>.
   \label{e:H-Derrick}
\end{equation}
Taking the first derivative of $H_{\mathrm{w}}$ with respect to $\beta$
we obtain
\begin{equation}\label{e:dHdbeta}
   \frac{1}{M} \pdv{H_{\mathrm{w}}(\beta)}{\beta} 
   = 
   \frac{1}{G_0} \, \Bigl [\, 1 - \frac{1}{\beta^2} \, \Bigr]
   + 
   g \, 
   \Big [\,
      \frac{\kappa}{(\beta + \kappa)^2}
      -
      \frac{\kappa \, \beta^{\kappa-1}}{(\kappa + 1)^2} \,
   \Big ] \,
   \Bigl [ \frac{M}{\pi \, G_0} \Bigr ]^{\kappa} \>.
\end{equation}
From Eq.~\eqref{e:dHdbeta}, we see that $\partial H/\partial \beta |_{\beta = 1} = 0$, 
therefore the solution we found is a stationary point of the stretched Hamiltonian. 
Taking the second derivative of $H_{\mathrm{w}}$ with respect to $\beta$, evaluating 
it at $\beta=1$ and dividing by the mass we obtain
\begin{equation}
   \frac{1}{M} \, \pddv{H_{\mathrm{w}}(\beta)}{\beta} \Big |_{\beta = 1} 
   = 
   \frac{2}{G_0} 
   -
   g \,
   \frac{\kappa(\kappa^2+1)}{(\kappa+1)^3} 
   \Bigl [\,
      \frac{M}{\pi G_0}\,
   \Bigr ]^{\kappa} \>.
   \label{Hwsecond}
\end{equation}
Derrick's theorem predicts that the soliton is stable to width 
perturbations (by keeping $M$ fixed), if Eq.~\eqref{Hwsecond} 
is positive, or
%
\begin{equation} 
   M < M_{\mathrm{w}}(\kappa) 
   = 
   \pi G_0 \,
   \Bigl [\,
      \frac{2}{g G_0} \, \frac{(\kappa+1)^3}{\kappa(\kappa^2+1)} \,
   \Bigr]^{1/\kappa} \>,
\end{equation}
which reduces into
%
\begin{equation}\label{e:Mstar}
   M < M_{\mathrm{w}}(\kappa)
   = 
   \pi \,
   \Bigl [\,
      \frac{2 \, (\kappa+1)^3}{\kappa \,(\kappa^2+1)} \,
   \Bigr ]^{1/\kappa} \>,
\end{equation}
upon setting $G_0 = 1$ (and $g=1$ as before). The behavior of critical 
mass $M^{\ast}_{\mathrm{w}}(\kappa)$ is shown in red in Fig.~\ref{fig:Mcrit}.
Since $M^{\ast} \rightarrow  \pi$ as $\kappa \rightarrow \infty$, the
exact solution is stable for all values of $\kappa$ provided that 
$M<\pi$.
In terms of the amplitude $\tilde{A}$ we have instead stability if 
\begin{equation}
   \tilde{A} < \tilde{A}_{\mathrm{w}}(\kappa) 
   = 
   \Bigl [\,
      \frac{2 \, (\kappa+1)^3}{\kappa \, (\kappa^2+1)} 
   \Bigr ]^{1/(2 \kappa)} \>.
\end{equation}
%
%
%
%

%
%
\subsection{\label{ss:TransDerrick}Translational stability}

Similar to Derrick's theorem for width stability, we can ask what happens 
when we shift the position of the solution away from the origin. For simplicity 
let us consider $x \rightarrow x+a$ and ask whether the energy of the solution 
goes up or down. We will find that $x=0$ is an extremum of the potential, and 
that there is a critical mass $M_\mathrm{t}$ which is dependent on $\kappa$, above which 
the exact solution becomes a maximum of $H(a,\kappa)$.  So we now consider 
the shifted wavefunction: 
\begin{equation}
   \tilde{\psi}(x,y,t) 
   =
   A_0 \, 
   \rme^{- [\, (x - a)^2 + y^2 \,]/(2 \, G_0) - \rmi \, \varphi(t) }
   \qc
   A_0^2 = \frac{M}{\pi G_0} \>.
\end{equation}
This shift in the position does not effect $H_1$ and $H_2$, and thus 
we get:
\begin{eqnarray}
   H_1 &=& \frac{M}{G_0} \>,
   \label{e:H1trans} \\
   H_2 &=&
   - 
   \frac{g \,M}{(\kappa + 1)^2} \, 
   \Bigl [ \frac{M}{\pi \, G_0} \Bigr ]^{\kappa} \>,
   \label{e:H2trans}  \\
   H_3
   &=&
   \int_{-\infty}^\infty \hspace{-1em} \dd{x} 
   \int_{-\infty}^\infty \hspace{-1em} \dd{y} 
   |\, \tilde{\psi}[x,y,Q(t)] \,|^2  \, V_1(x,y)
   \notag \\
   &=&   
   g \, \Bigl ( \frac{M }{\pi G_0} \Bigr )^{\kappa+1}
   \int_{-\infty}^\infty \hspace{-1em} \dd{x} 
   \int_{-\infty}^\infty \hspace{-1em} \dd{y} \,
   \rme^{ - [(x - a)^2 + y^2 + \kappa (x^2 + y^2) ]/G_0 }
   \label{e:H3trans}  \\
   &=&
   \frac{g \, M }{\kappa + 1 } \,
   \Bigl [ \frac{M}{\pi \, G_0} \Bigr ]^{\kappa} \,
   \rme^{- \kappa \, a^2 / [\, (\kappa + 1) \, G_0 \,] } \>,
   \notag \\
   H_4 
   &=& 
   \int_{-\infty}^\infty \hspace{-1em} \dd{x} 
   \int_{-\infty}^\infty \hspace{-1em} \dd{y} 
   |\, \tilde{\psi}[x,y,Q(t)] \,|^2  \, V_2(x,y)
   \notag \\
   &=&
   \frac{M }{\pi G_0^3}
   \int_{-\infty}^\infty \hspace{-1em} \dd{x} 
   \int_{-\infty}^\infty \hspace{-1em} \dd{y} \,
   ( x^2 + y^2 ) \,
   \rme^{- [\, (x - a)^2 + y^2 \,]/G_0}
   =\frac{M}{G_0^2} \, (\, G_0 + a^2 \,) \>.
   \label{e:H4trans} 
\end{eqnarray}
This way, the displaced Hamiltonian denoted by $H_{\mathrm{t}}(a)$
reads
\begin{eqnarray}
   \frac{H_{\mathrm{t}}(a)}{M}
   &=&
   \frac{1}{G_0}
   +
   \frac{g}{(\kappa + 1)^2} \, 
   \Bigl [ \frac{M}{\pi \, G_0} \Bigr ]^{\kappa}
   \label{e:Htrans} \\
   && \hspace{1em}
   +
   \frac{g}{\kappa + 1 } \,
   \Bigl [ \frac{M}{\pi \, G_0} \Bigr ]^{\kappa} \,
   \rme^{- \kappa \, a^2 / [\, (\kappa + 1) \, G_0 \,] }
   +
   \frac{1}{G_0^2} \, (\, G_0 + a^2 \,) \>.
   \notag
\end{eqnarray}
The first derivative of this expression with respect to $a$ is 
\begin{equation}\label{e:dHtransda}
   \pdv{H_{\mathrm{t}}(a)}{a}
   =
   -
   \frac{2 g \, \kappa \, a}{(\kappa + 1)^2 } \,
   \Bigl [ \frac{M}{\pi \, G_0} \Bigr ]^{\kappa} \,
   \rme^{- \kappa \, a^2 / [\, (\kappa + 1) \, G_0 \,] }
   +
   \frac{2 \, a}{G_0^2} \>,
\end{equation}
and gives zero at $a=0$, showing that the exact solution is indeed an 
extremum of the energy.  The second derivative at $a=0$ yields
\begin{equation}
   \pddv{H_{\mathrm{t}}(a)}{a} \Big |_{a=0} 
   = 
   \frac{2}{G_0^2} 
   -
   \frac{2 g \, \kappa}{(\kappa + 1)^2 } \,
   \Bigl [ \frac{M}{\pi \, G_0} \Bigr ]^{\kappa} \>,
\end{equation}
and stability with respect to translations $a$ (again, 
while keeping $M$ fixed), requires that 
\begin{equation}\label{MtFull} 
   M
   <
   M_{\mathrm{t}}(\kappa)
   =
   \pi G_0 \Bigl [ \frac{(\kappa+1)^2}{g \, G_0^2 \, \kappa} \Bigr ]^{1/\kappa} \>,
\end{equation}
which reduces into
%
\begin{equation}\label{e:Mt} 
   M
   <
   M_{\mathrm{t}}(\kappa)
   =
   \pi \Bigl [ \frac{(\kappa+1)^2}{\kappa} \Bigr ]^{1/\kappa} \>,
\end{equation}
upon setting $G_0 = 1$ (and $g=1$ again). We see that $M_{\mathrm{t}}^{\ast}(\kappa) > \pi$, 
so that as long as $M < \pi$ there is no translational instability.  The curve for 
$M_{\mathrm{t}}^{\ast}(\kappa)$ is shown in red in Fig.~\ref{fig:Mcrit} and compared to 
the critical mass for the width instability. By comparing \eqref{e:Mt} with \eqref{e:Mstar}, 
we find that there is a crossover effect at $\kappa = 1 + \sqrt{2}$. Below $\kappa = 1 + \sqrt{2}$, 
the translational instability occurs first.  Above this value the width instability occurs first.
It is worth pointing out again that when $M < \pi$, there is neither translational nor width 
instability regardless of the value of $\kappa$.
%
%
\begin{figure*}[t]
   \centering
   \subfigure[\ Stable case: $M=(4/5)\, M_{\mathrm{t}}$.]
   {\label{k1-2stable}
   \includegraphics[width=0.4\linewidth]{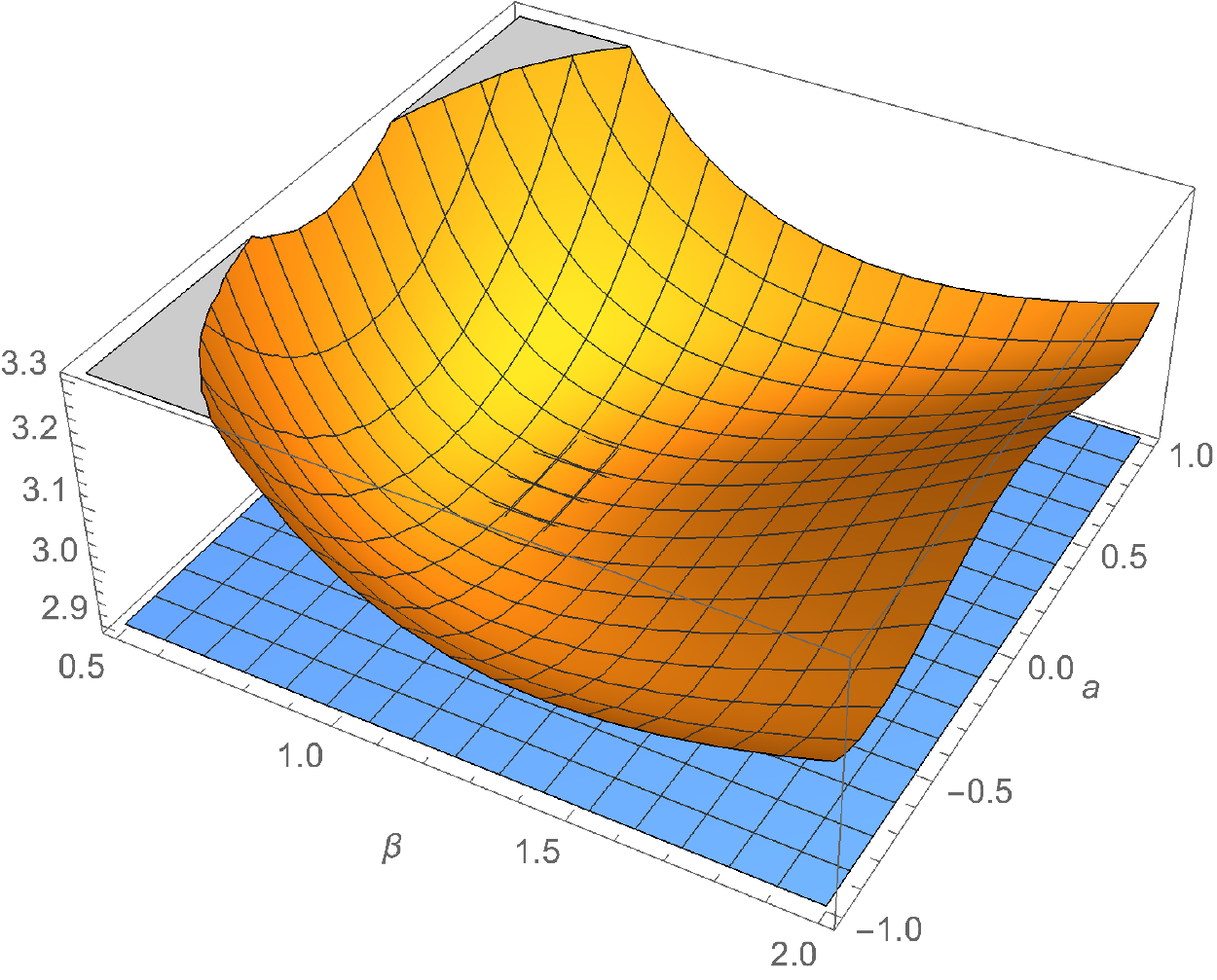} }
   \hspace{0.5em}
   \subfigure[\ Unstable case: $M=(11/10)\, M_{\mathrm{w}}$.]
   {\label{k1-2unstable}
   \includegraphics[width=0.4\linewidth]{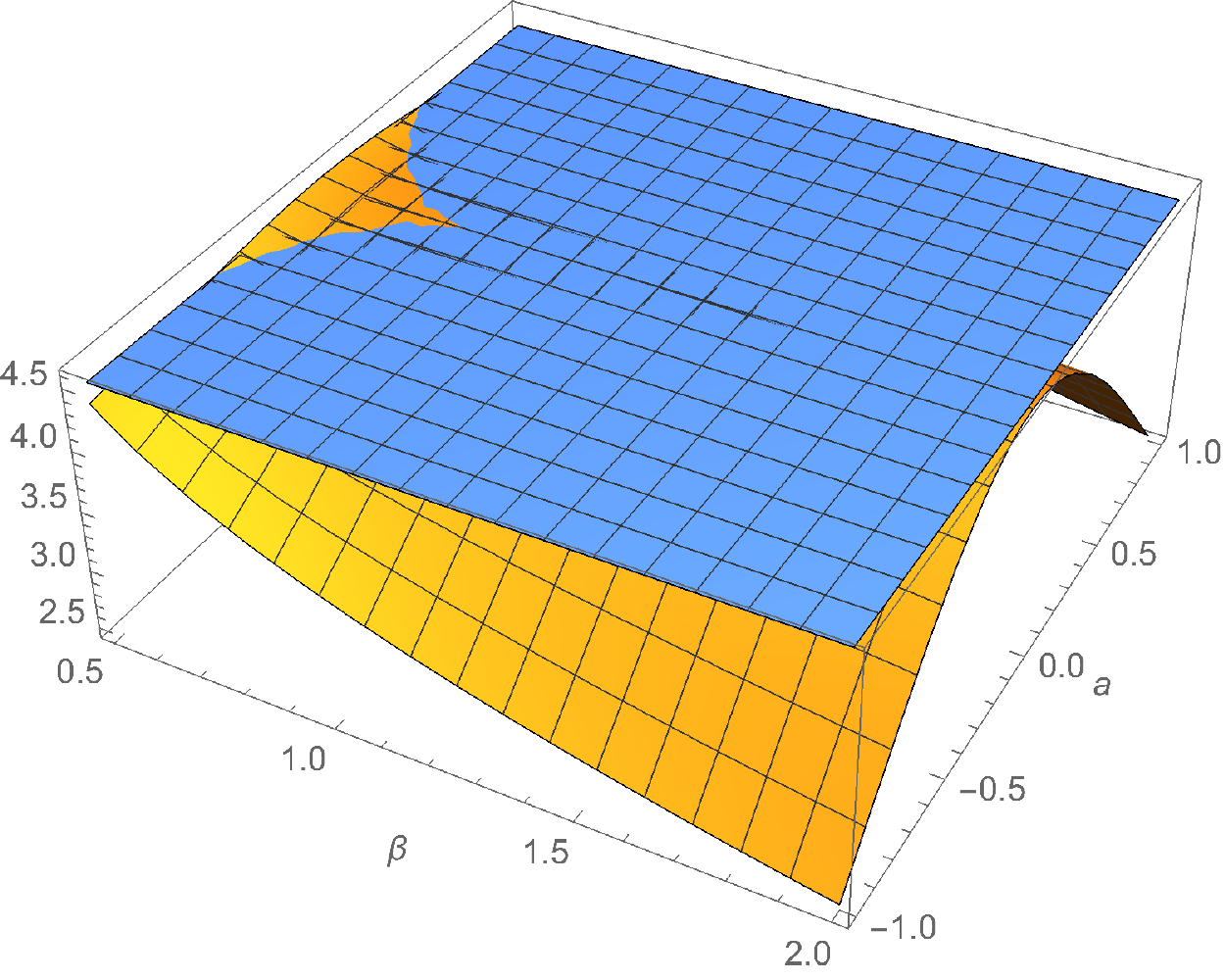} }
   \caption{\label{fig:EnergyLand}Plots of the energy landscapes 
   $E(a,\beta)/M$ (yellow) and the exact energy 
   $E(0,1)/M$ (blue) for $\kappa = 1/2$ for two values of $M$.}
\end{figure*}
%
%
\subsection{\label{se:PotentialLandscape}The potential energy landscape}

Stability for both translations and stretches can be studied through the 
wavefunction of the form of
\begin{equation}\label{e:TWwavefct}
   \tilde{\psi}(x,y,t) 
   =
   A \, \rme^{-\beta [\,(x - a)^2 + y^2 \,]/2 - \rmi \, \varphi(t) }
   \qc
   A^2 = \frac{\beta \, M}{\pi} = \beta \, A_0^2 \>,
\end{equation}
%
whose total energy is given by
\begin{equation}\label{e:Halphabeta}
   \frac{E(a,\beta)}{M}
   =
   a^2
   +
   \beta
   +
   \frac{1}{\beta}
   -
   \Bigl [ \, \frac{M}{\pi} \Bigr ]^{\kappa} \,
   \Bigl [ \,
      \frac{\beta^{\kappa}}{(\kappa+1)^2} 
      -
      \frac{\beta}{\kappa + \beta} \, 
      \rme^{ - \kappa \, \beta \, a^2 / (\kappa + \beta) } \,
   \Bigr ] \>.
\end{equation}
There are two critical masses for translational and width instabilities,
given respectively by
\begin{eqnarray}
   M_{\mathrm{t}}
   &=& 
   \pi \, [(\kappa +1)^2/\kappa]^{1/\kappa } \>,
   \label{e:MtransX} \\
   M_{\mathrm{w}} 
   &=&
   \pi \, [\, 2 \, (\kappa+1)^3/( \kappa (\kappa^2+1) \,) \,]^{1/\kappa} \>.
   \label{e:MwidthX}
\end{eqnarray}
For the exact energy, $\alpha = 0$ and $\beta = 1$,
\begin{equation}\label{e:Ealphabeta00}
   \frac{E(0,1)}{M}
   =
   2 + \frac{\kappa}{(\kappa+1)^2} \, \Bigl [ \, \frac{M}{\pi} \Bigr ]^{\kappa} \>,
\end{equation}
which is in agreement with Eq.~\eqref{e:EnergyExact}.
   To show how intricate the energy landscape can be, we display two cases for $\kappa=1/2$ in Fig.~\ref{fig:EnergyLand}.  If we are in the regime where the mass is less than both critical masses, then by choosing $M = (4/5) \, M_{\mathrm{t}}$, we get the results shown in Fig.~\ref{k1-2stable}.
If instead we choose $M = (11/10) \, M_{\mathrm{w}}$, then we are in the unstable regime as shown in Fig.~\ref{k1-2unstable} . 

%
%
\subsection{\label{ss:DerrickNoPot}Derrick's theorem in the absence of a potential}

In contrast, when $V(\mathbf{r})\equiv 0$, Derrick's theorem for width stability 
does not provide one with a critical mass.  
Instead, from Eq.~\eqref{e:H-Derrick} (and for $g = G_0 = 1$), we directly obtain
\begin{equation}
   \frac{H(\beta)}{M}
   = 
   \beta
   -
   \frac{1}{(\kappa + 1)^2} \, 
   \Bigl [ \frac{\beta M}{\pi} \Bigr ]^{\kappa} \>,
\end{equation}
%
%
whose first derivative yields
\begin{equation}\label{e:V0dHdbeta} 
   \frac{1}{M} \pdv{H}{\beta} 
   = 
   1
   -
   \frac{\kappa \, \beta^{\kappa - 1}}{(\kappa + 1)^2} \, 
   \Bigl [ \frac{M}{\pi} \Bigr ]^{\kappa} \>.
\end{equation}
Choosing $M = M_1$, where
\begin{equation}\label{e:M1value-I}
   M_1
   =
   \pi \, \Bigl [ \frac{(\kappa+1)^2 }{\kappa} \Bigr ]^{1/\kappa} \>,
\end{equation}
then Eq.~\eqref{e:V0dHdbeta} vanishes at $\beta = 1$ showing that 
this is an extremum. The condition for this to be a minimum is that 
\begin{equation}
   \frac{1}{M}
   \pddv{H}{\beta} \Big |_{\beta=1} 
   =
   -  
   \frac{\kappa \, (\kappa - 1 )}
        {(\kappa +1)^2}
   \Bigl [ \frac{M_1}{\pi}\Bigr ]^{\kappa}
   =
   1 - \kappa > 0 \>,
\end{equation}
so that for the 2D NLSE, stability is guaranteed as long as $\kappa < 1$. 
In arbitrary dimensions $d$ a similar calculation yields stability for 
$\kappa d < 2$.

%
%
\section{\label{s:CCmethod}Collective coordinate method}

The collective coordinate (CC) method uses a variational \ansatz\ to 
solve for the dynamics from the action given in Eq.~\eqref{GammaDef} 
for the NLSE in an external potential. In this paper we will employ a four 
CC \ansatz\ so that we can explore the response of the solution when we 
perturb the initial wavefunction both in the width as well as in the position. 
The method we use here is similar to the method introduced in a previous paper,
and authored by some of the current authors \cite{Dawson-2017}.
We restrict our calculation here to 4CCs, which allows us to recover the 
results of Derrick's theorem. However, by comparing these with numerical 
results of the 
NLSE in the unstable regime, we find that translations in the $y$ direction, 
which were not included here, get excited. Also, once instabilities manifest themselves, 
the shape of the wavefunction starts deviating from our assumed Gaussian shape.  

\subsection{\label{ss:2CCmethod}Two collective coordinate (2CC) ansatz}

If we are just interested in the dynamics of the width of self-similar solutions,  
we can assume that the wavefunction can be parametrized by two CCs, and 
thus choose
\begin{eqnarray}\label{e:2CCansatz}
   \tilde{\psi}[r,Q(t)]
   &=&  
   \tilde{A}(t) \, 
   \rme^{- r^2/[2 G(t)] + \rmi \, \Lambda (t) \, r^2 - \rmi \, \varphi(t) }\>,
   \notag \\
   M 
   &=& 
   \int \ddd{x} |\tilde{\psi}[r,Q(t)]|^2
   = 
   \pi \, G(t) \, \tilde{A}^2(t) \>,
   \notag \\
   Q(t) 
   &=& 
   \{ G(t), \Lambda (t) \} \>.
\end{eqnarray}
Here $\tilde{A}(t)$ is fixed by $M$ and $G(t)$ so $\varphi(t)$ is irrelevant 
to the dynamics. This Gaussian \ansatz\ \eqref{e:2CCansatz} agrees with 
the results of Perez-Garcia~\cite{Perez-Garcia-2004}, who showed that if 
one has a self-similar solution of  the NLSE of the form 
\begin{equation}
   \tilde{\psi}[r,w(t),\phi(r,t)] 
   = 
   \tilde{A}(t) \, \rho \Bigl [\frac{r}{w(t)} \Bigr ] \, \rme^{\rmi \, \phi(r,t)}\>,
\end{equation}
then the phase is fixed to be quadratic and of the form
\begin{equation}
   \phi(r,t) = \frac{\dot w}{2w} \, r^2 \>.
\end{equation}
From Lagrange's equations for the collective coordinates (see below) we will find 
\begin{equation}
\Lambda= \dot G/(8G) \>.
\end{equation}

\subsection{\label{ss:4CCmethod} Four collective coordinate (4CC) ansatz}
To compare with our energy landscape static calculation above, it is sufficient 
to consider the response of the wavefunction to translations in one spatial 
direction, which we will choose to be the $x$ direction. Indeed, we can 
study the response of the wavefunction to small perturbations in width and position
through a suitable 4CC \ansatz\ in a variational approach by replacing 
\begin{equation}
   x^2 + y^2 \rightarrow \bar{x}^2(t) + y^2
   \qc
   \bar{x}(t) = x - q_x(t)\>.
\end{equation}
The conjugate coordinate to $q_x(t)$ is the momentum $p_x(t)$ as a collective 
coordinate.  For simplicity, we will suppress the subindex $x$ on $q,p$, and 
choose for our 4CC variational wavefunction:
\begin{eqnarray}
\fl
   \tilde{\psi}[x,y,Q(t)] 
   &=& 
   A(t) \, \rme^{\phi(x,y,t)}
   \qc
   M
   =
   \pi \, G(t) \, A^2(t)
   =
   \pi \, G_0 \, A_0^2 \>,
   \notag \\
\fl
   \phi[x,y,Q(t)]
   &=&
   -
   \frac{\bar{x}^2(t) + y^2}{2 G(t)}
   + 
   \rmi \,
   [\,
      p(t) \, \bar{x}(t)
      + 
      \Lambda(t) \, (\bar{x}^2(t) + y^2)
      +
      \varphi(t) \,
    ]  \>.
   \label{e:4CCpsi}
\end{eqnarray}
Here again $A(t)$ is fixed by $M$ and $G(t)$ and is not a dynamic variable. 
This means that $\varphi(t)$ is not dynamic either, and we ignore it in the 
following, so then the four generalized coordinates are: $Q(t) = \{\, q(t),p(t),G(t),\Lambda(t) \,\}$.
The $x$-displacement $q(t)$ and width $G(t)$ are then given by the integrals:
\begin{eqnarray}
   q(t)
   &=&
   \frac{1}{M}
   \int_{-\infty}^\infty \hspace{-1em} \dd{x} 
   \int_{-\infty}^\infty \hspace{-1em} \dd{y} 
   |\, \tilde{\psi}[x,y,Q(t)] \,|^2 \, x\>,
   \label{e:Mqint}
   \\
   G(t)
   &=&
   \frac{1}{M}
   \int_{-\infty}^\infty \hspace{-1em} \dd{x} 
   \int_{-\infty}^\infty \hspace{-1em} \dd{y} 
   |\, \tilde{\psi}[x,y,Q(t)] \,|^2 \, [\, \bar{x}^2(t) + y^2 \,]
   \label{e:MGint} \>.
\end{eqnarray}
Using Eqs.~\eref{e:Mqint} and~\eqref{e:MGint}, it is easy to extract 
the variational parameters from simulations by calculating the first 
two moments of the density. When we insert the variational wavefunction 
into the complete action of Eq.~\eqref{GammaDef} and integrate over the 
spatial degrees of freedom, we get an effective action for the variational 
parameters. In this process, we keep the parameters of the potential fixed 
by the exact solution. Writing the external potential in terms of the conserved 
mass $M$ with $V(r) = V_1(r) + V_2(r)$, from Eq.~\eqref{Vassume1}, we have
\begin{equation}
   V_1(r) 
   = 
   g \, \Bigl (\frac{M}{ \pi G_0} \Bigr )^\kappa \rme^{- \kappa \, r^2/G_0}
   \qc
   V_2(r) 
   = 
   \Bigl ( \frac{r}{G_0} \Bigr )^2 \>.
\end{equation}
The action then takes the form
\begin{equation}
   \Gamma[Q] = \tint \dd{t} L[\,Q,\dot{Q}]\>,
\end{equation}
where the Lagrangian is given by
\begin{equation}
   L[\,Q,\dot{Q}]
   = 
   M \, [\, p(t) \, \dot{q}(t) + \Lambda(t) \, \dot{G}(t) \,] - H[\,Q\,] \>.
\end{equation}
The Hamiltonian is a sum of four terms
\begin{equation}\label{e:Hfourterms}
   H(Q) = H_1(Q) + H_2(Q) + H_3(Q) + H_4(Q) \>,
\end{equation}
where
\begin{eqnarray}
   H_1(Q)
   &=&
   \int_{-\infty}^\infty \hspace{-1em} \dd{x} 
   \int_{-\infty}^\infty \hspace{-1em} \dd{y} 
   \Bigl [\, 
      |\partial_x \tilde{\psi}(x,y) |^2 + |\partial_y \tilde{\psi}(x,y) |^2 \,
   \Bigr ] \>,
   \notag \\
   &=&
   \int_{-\infty}^\infty \hspace{-1em} \dd{x} 
   \int_{-\infty}^\infty \hspace{-1em} \dd{y} 
   \Bigl [\, 
      | \bar{x}/G + \rmi p + 2 \rmi \Lambda \bar{x} |^2
      +
      |  y/G +  2 \rmi \Lambda y |^2 \,
   \Bigr ] \, | \tilde{\psi} |^2 \>,
   \notag \\
   &=&
   \frac{M}{\pi G}
   \int_{-\infty}^\infty \hspace{-1em} \dd{x} 
   \int_{-\infty}^\infty \hspace{-1em} \dd{y} \,
   \rme^{-(\bar{x}^2+y^2)/G} \,
   \Bigl [\,
      p^2 + 4 \,\Lambda^2 (\,\bar{x}^2 + y^2) + \frac{\bar{x}^2+y^2}{G^2} \,
   \Bigr ]
   \notag \\
   &=&
   M \, ( p^2 + 4 \, G \, \Lambda^2 + 1/G) \>,
   \label{e:4CC-H1} \\
   H_2(Q)
   &=&
   - \frac{g}{\kappa+1}
   \int_{-\infty}^\infty \hspace{-1em} \dd{x} 
   \int_{-\infty}^\infty \hspace{-1em} \dd{y} \,
   | \tilde{\psi}(x,y) |^{2 \kappa + 2}
   \notag \\
   &=&
   - \frac{g}{\kappa+1} \, \Bigl [ \, \frac{M}{\pi G} \Bigr ]^{\kappa + 1}
   \int_{-\infty}^\infty \hspace{-1em} \dd{x} 
   \int_{-\infty}^\infty \hspace{-1em} \dd{y} \,
   \rme^{- (\kappa + 1) ( \bar{x}^2 + y^2 )/G} 
   \notag \\
   &=&
   - \frac{g M}{(\kappa+1)^2} \, \Bigl [ \, \frac{M}{\pi G} \Bigr ]^{\kappa} \>,
   \label{e:4CC-H2} \\
   H_3(Q)
   &=&
   \int_{-\infty}^\infty \hspace{-1em} \dd{x} 
   \int_{-\infty}^\infty \hspace{-1em} \dd{y} \,
   | \tilde{\psi}(x,y) |^{2} \,
   V_1(x,y)
   \notag \\
   &=&
   g \, \Bigl [ \, \frac{M}{\pi G_0} \Bigr ]^{\kappa} \,
   \Bigl [ \, \frac{M}{\pi G} \Bigr ] \,
   \int_{-\infty}^\infty \hspace{-1em} \dd{x} 
   \int_{-\infty}^\infty \hspace{-1em} \dd{y} \,
   \rme^{- \kappa (x^2 + y^2)/G_0 - (\bar{x}^2 + y^2)/G }
   \notag \\
   &=&
   g \, \frac{M \, G_0}{G \kappa + G_0} \, \Bigl [ \, \frac{M}{\pi G_0} \Bigr ]^{\kappa} \,
   \rme^{ - \kappa \, q^2 / (\kappa G + G_0)  } \>,
   \label{e:4CC-H3} \\
   H_4(Q)
   &=&
   \int_{-\infty}^\infty \hspace{-1em} \dd{x} 
   \int_{-\infty}^\infty \hspace{-1em} \dd{y} \,
   | \tilde{\psi}(x,y) |^{2} \,
   V_2(x,y)
   \notag \\
   &=&
   \frac{1}{G_0^2} \, \Bigl [ \, \frac{M}{\pi G} \Bigr ]
   \int_{-\infty}^\infty \hspace{-1em} \dd{x} 
   \int_{-\infty}^\infty \hspace{-1em} \dd{y} \,
   (\, x^2 + y^2 \,) \,
   \rme^{ - (\bar{x}^2 + y^2)/G }
   \notag \\
   &=&
   \frac{M}{G_0^2} \, (\, G + q^2 \,) \>.
   \label{e:4CC-H4}
\end{eqnarray}
Adding these terms, the total Hamiltonian is given by
\begin{eqnarray}
\fl
   \frac{H(Q)}{M}
   &=&
   p^2
   + 
   \frac{q^2}{G_0^2}
   +
   \frac{G}{G_0^2}
   +
   \frac{1}{G} 
   + 
   4 \, G \, \Lambda^2 
   -
   \frac{g}{(\kappa+1)^2} \, \Bigl [ \, \frac{M}{\pi G} \Bigr ]^{\kappa}
   \notag \\
\fl
   &&\hspace{2em}
   +
   \frac{g \, G_0}{G \kappa + G_0} \, \Bigl [ \, \frac{M}{\pi G_0} \Bigr ]^{\kappa} \,
   \rme^{ - \kappa \, q^2 / (\kappa G + G_0) } \>.
   \label{e:4CC-Hamiltonian}
\end{eqnarray}
Note that Eq.~\eqref{e:4CC-Hamiltonian} agrees with Eq.~\eqref{e:EnergyExact} when 
$q = p = \Lambda = 0$. The Lagrangian for the 4CC \ansatz\ is then given by:
\begin{eqnarray}
\fl
   \frac{L[\,Q,\dot{Q}\,]}{M}
   &=&
   p \, \dot{q} + \Lambda \, \dot{G}
   -
   p^2
   - 
   \frac{q^2}{G_0^2}
   -
   \frac{G}{G_0^2}
   -
   \frac{1}{G} 
   - 
   4 \, G \, \Lambda^2 
   +
   \frac{g}{(\kappa+1)^2} \, \Bigl [ \, \frac{M}{\pi G} \Bigr ]^{\kappa}
   \notag \\
\fl
   &&\hspace{2em}
   -
   \frac{g \, G_0}{G\kappa + G_0} \, \Bigl [ \, \frac{M}{\pi G_0} \Bigr ]^{\kappa} \,
   \rme^{ - \kappa \, q^2 / (\kappa G + G_0) } \>.
   \label{e:4CC-Lagrangian}
\end{eqnarray}
From Eq.~\eqref{e:4CC-Lagrangian}, the equations of motion are
\begin{eqnarray}\label{e:4CC-EOM}
   \dot{q} 
   &=& 
   2 \, p \>,
   \label{e:4CC-EOM-a} 
   \\
   \dot{p}
   &=&
   - 
   \frac{2 \,q}{G_0^2}
   +
   \frac{2g \, \kappa \, G_0 \, q}{(G \kappa + G_0)^2} \, 
   \Bigl [ \, \frac{M}{\pi G_0} \Bigr ]^{\kappa} \,
   \rme^{ - \kappa \, q^2 / (G \kappa + G_0) } \>,
   \label{e:4CC-EOM-b} \\
   \dot{G}
   &=&
   8 \, G \, \Lambda \>,
   \label{e:4CC-EOM-c} \\
   \dot{\Lambda}
   &=&
   - 
   4 \, \Lambda^2
   -
   \frac{1}{G_0^2}
   +
   \frac{1}{G^2}
   -
   \frac{g \, \kappa}{(\kappa + 1)^2 \, G} \, \Bigl [ \, \frac{M}{\pi G} \Bigr ]^{\kappa}
   \notag \\
   && \hspace{1em}
   +
   \frac{g \, G_0 \, \kappa}{(G\kappa + G_0)^2} \,
   \Bigl [\,
      1 - \frac{\kappa \,q^2}{G\kappa + G_0} \,
   \Bigr ] \,
   \Bigl [ \, \frac{M}{\pi G_0} \Bigr ]^{\kappa} \,
   \rme^{ - \kappa \, q^2 /(G \kappa + G_0)} \>.
   \label{e:4CC-EOM-d}
\end{eqnarray}

%
%
\subsection{\label{e:NoPotential}Blowup time}

Using the equation of motion for $\dot G$, and setting $G_0 = 1$, 
we can rewrite the energy as
\begin{eqnarray} \label{energycc} 
   \frac{E(Q)}{M}
   &=&
   p^2
   + 
   q^2
   + 
   \frac{\dot{G}^2}{16 \, G}
   +
   G 
   + 
   \frac{1}{G}
   -
   \frac{g}{(\kappa+1)^2} \, \Bigl [ \, \frac{M}{\pi\, G} \Bigr ]^{\kappa}
   \notag \\
   && \hspace{1em} 
   +
   \frac{g}{G \kappa + 1} \, \Bigl [ \, \frac{M}{\pi} \Bigr ]^{\kappa} \,
   \rme^{ - \kappa \, q^2 / (\kappa G + 1) } \>.
\end{eqnarray}
We will see below from our simulations that one can have  blowup ($G \rightarrow 0$), 
as long as  $\kappa \ge 2$ and $M > M^\ast $.  The energy is conserved, and 
constrains the range of $G$ and $q$.  The initial energy of the perturbed solution is given 
by Eq.~\eqref{energycc} with $q=q(t=0), p=p(t=0), G=G(t=0), \dot G= \dot G (t=0)$, which 
for our simulations will be close to the energy of the exact solution $E=E_0$, and is given 
by Eq.~\eqref{e:EnergyExact}, or
\begin{equation}
   \frac{E_0}{M} 
   = 
   2
   +
   \frac{g \, \kappa}{(\kappa + 1)^2} \, \Bigl [ \frac{M}{\pi} \Bigr ]^{\kappa}\>.
   \label{e:EnergyExactII}
\end{equation}
When $G \rightarrow 0$, from the leading terms (that must cancel), 
we obtain
%
\begin{equation}
   \dot{G}
   = 
   - \sqrt{\frac{16\, G(t)}{(\kappa +1)^2} \, \Bigl[ \frac{M}{\pi \, G(t)} \Bigr]^{\kappa} }
   =
   - \frac{4 M^{\kappa/2}} {\pi^{\kappa/2} (\kappa+1)} G(t) ^{(1-\kappa)/2} \>,
\end{equation}
which can be integrated, thus yielding (near the blowup time $t^{\ast}$ with $t<t^{\ast}$)
%
%
\begin{equation}
   G(t) 
   = 
   \frac{4 M^{\kappa/2}}{\pi^{\kappa/2} (\kappa+1)^2} \, (t-t^{\ast}) ^{2/(\kappa+1)} \>.  
\end{equation}
References to blowup times can be found in \cite{Ball-1977,Glassey-1977}.

%
%
\subsection{\label{ss:SmallAmp4CC}Small amplitude approximation for the 4CC dynamics} 

From Eqs.~(\ref{e:4CC-EOM-a} - \ref{e:4CC-EOM-d}), we can obtain small 
oscillation equations by setting $G_0 = 1$, letting
\begin{equation}\label{e:SOset}
   q = \delta q 
   \qc
   p = \delta p
   \qc
   G = 1 + \delta G
   \qc
   \Lambda = \delta \Lambda  \qc \left(\delta q, \delta p, \delta G, \delta \Lambda \ll 1\right)\>,
\end{equation}
and keeping only the linear terms.  We obtain:
\begin{eqnarray}\label{e:}
   \delta \dot{q}
   &=&
   2 \, \delta p \>,
   \label{e:SO-a} \\
   \delta \dot{p}
   &=&
   - 
   2 \, 
   \Bigl \{ \,
      1
      -
      \frac{g  \kappa}{(\kappa+1)^2} \, 
   \Bigl [ \, \frac{M}{\pi} \Bigr ]^{\kappa} \,
   \Bigr \} \, \delta q \>,
   \label{e:SO-b} \\
   \delta \dot{G}
   &=&
   8 \, \delta \Lambda \>,
   \label{e:SO-c} \\
   \delta \dot{\Lambda}
   &=&
   -
   \Bigl \{\,
      2
      -
      \frac{g \, \kappa(\kappa^2 + 1)}{(\kappa+1)^3} \,
      \Bigl [ \, \frac{M}{\pi} \Bigr ]^{\kappa} \,
   \Bigr \} \, \delta G \>. 
   \label{e:SO-d}
\end{eqnarray}
We observe from the above that the $(\delta q, \delta p)$ dynamics 
decouple from the $(\delta G, \delta \Lambda)$ dynamics, and thus 
we find the small oscillations are governed by the equations
\begin{equation}\label{e:SOdynamics}
   \delta \ddot{q} + \omega_q^2 \, \delta q = 0
   \qc
   \delta \ddot{G} + \omega_G^2 \, \delta G = 0 \>,
\end{equation}
with
\begin{eqnarray}\label{e:omegaq-omegaG}
   \omega_q^2
   &=&
   4 \, 
      \Bigl \{ \,
      1
      -
      \frac{g \, \kappa}{(\kappa+1)^2} \, 
      \Bigl [ \, \frac{M}{\pi} \Bigr ]^{\kappa} \,
   \Bigr \}
   =
   4 \, \Bigl \{\, 
     1 - \Bigl [ \, \frac{M}{M_\mathrm{t}} \Bigr ]^{\kappa} \,
   \Bigr \} \>,
   \\
   \omega_G^2
   &=&
   8 \,
   \Bigl \{\,
      2
      -
      \frac{g \, \kappa \, (\kappa^2 + 1)}{(\kappa+1)^3} \,
      \Bigl [ \, \frac{M}{\pi} \Bigr ]^{\kappa} \,
   \Bigr \}
   =
   16 \, \Bigl \{\,
        1 - \Bigl [ \, \frac{M}{M_\mathrm{w}} \Bigr ]^{\kappa} \,
   \Bigr \} \>,
\end{eqnarray}
where $M_\mathrm{t}$ and $M_\mathrm{w}$ are given in Eqs.~(\ref{e:MtransX}-\ref{e:MwidthX}).  
For the $\delta q$ dynamics (translational) to be stable, we must have 
$M < M_\mathrm{t}$, and for the $\delta G$ dynamics (width) to be stable, 
we must have $M < M_{\mathrm{w}}$.
 
%
%
\section{\label{s:4CCevolutions}Typical evolutions in the 4CC approximation}
 
Here we explore the behavior of the 4CC ansatz for $\kappa$ in the range 
$1/2 \leq \kappa \leq 3/2$ which surrounds the critical value of $\kappa=1$
for blowup in the absence of a potential.  We consider three cases, 
$\kappa = 1/2$, $\kappa = 1$, and $\kappa = 3/2$.  For these three cases 
we choose masses in three regimes:
\begin{enumerate}[{Case} (a)]
   \item $M < M_{\mathrm{t}}, M_{\mathrm{w}}$ \,, 
   \item $M_{\mathrm{t}} < M < M_{\mathrm{w}}$ \,, 
   \item  $M_{\mathrm{t}},M_{\mathrm{w}} < M$ \,. 
\end{enumerate}
For illustrative purposes for the 4CC simulations, we will take for the exact 
solution: $g=G_0=1$, and the  initial values of $G(t=0) = 0.99$, $\Lambda(t=0) = 0$, 
$q(t=0) =0.01$, and $p(t=0) = 0$.  The values of $M$, initial values of $A_0$ 
and the energy $E(t=0) $ for the initial trial wavefunction of Eq.~\eqref{e:4CCpsi} 
are given in Table~\ref{t:plotdata}.

%
%
\begin{table}
\centering
\caption{Values of $\kappa$, mass ($M$), and energy ($E$) used in 
Section~\ref{s:4CCevolutions} and plotted in Figure \ref{fig:Mcrit-a}. 
We use initial values of $G_0 = 0.99$, $\Lambda_0 = 0$, $q_0 =0.01$, and $p_0 = 0$.}
\label{t:plotdata}
\footnotesize
\begin{tabular}{@{}lcccc}
\br
$\kappa$&case&$M$&$A_0$&$E_0$\\
\mr
$1/2$&(a)&50&4.0095&2.8866\\
$$&(b)&175 &7.5011&3.6585\\
$$&(c)&419&11.6069&4.5662\\
$1$&(a)&11&1.8806&2.8754\\
$$&(b)&17&2.3379&3.3528\\
$$&(c)&28&3.0005&4.2280\\
$3/2$&(a)&7&1.5002&2.7983\\
$$&(b)&9.4&1.7385&3.2422\\
$$&(c)&12&1.9642&3.7916\\
\br
\end{tabular}
\end{table}
%
%

%
\subsection{$\kappa = 1/2$}

\begin{figure}[t]
   \centering
   \subfigure[$q(t)$ case a]
   {\label{f:q-Kappa-05-Case-a}
   \includegraphics[width=0.475\columnwidth]{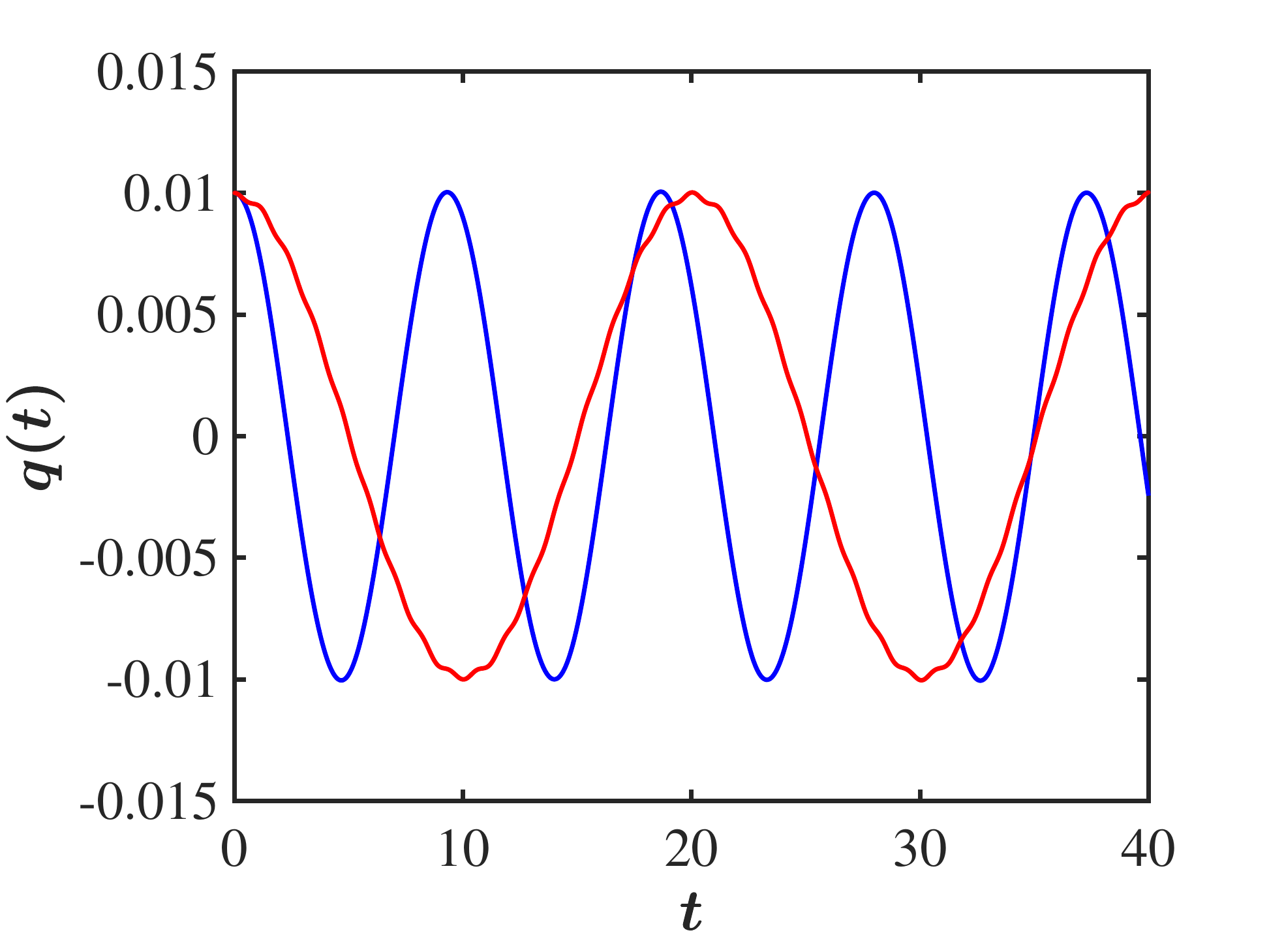}}
   \hspace{0.5em}
   \subfigure[$G(t)$ case a]
   {\label{f:G-Kappa-05-Case-a}
   \includegraphics[width=0.475\columnwidth]{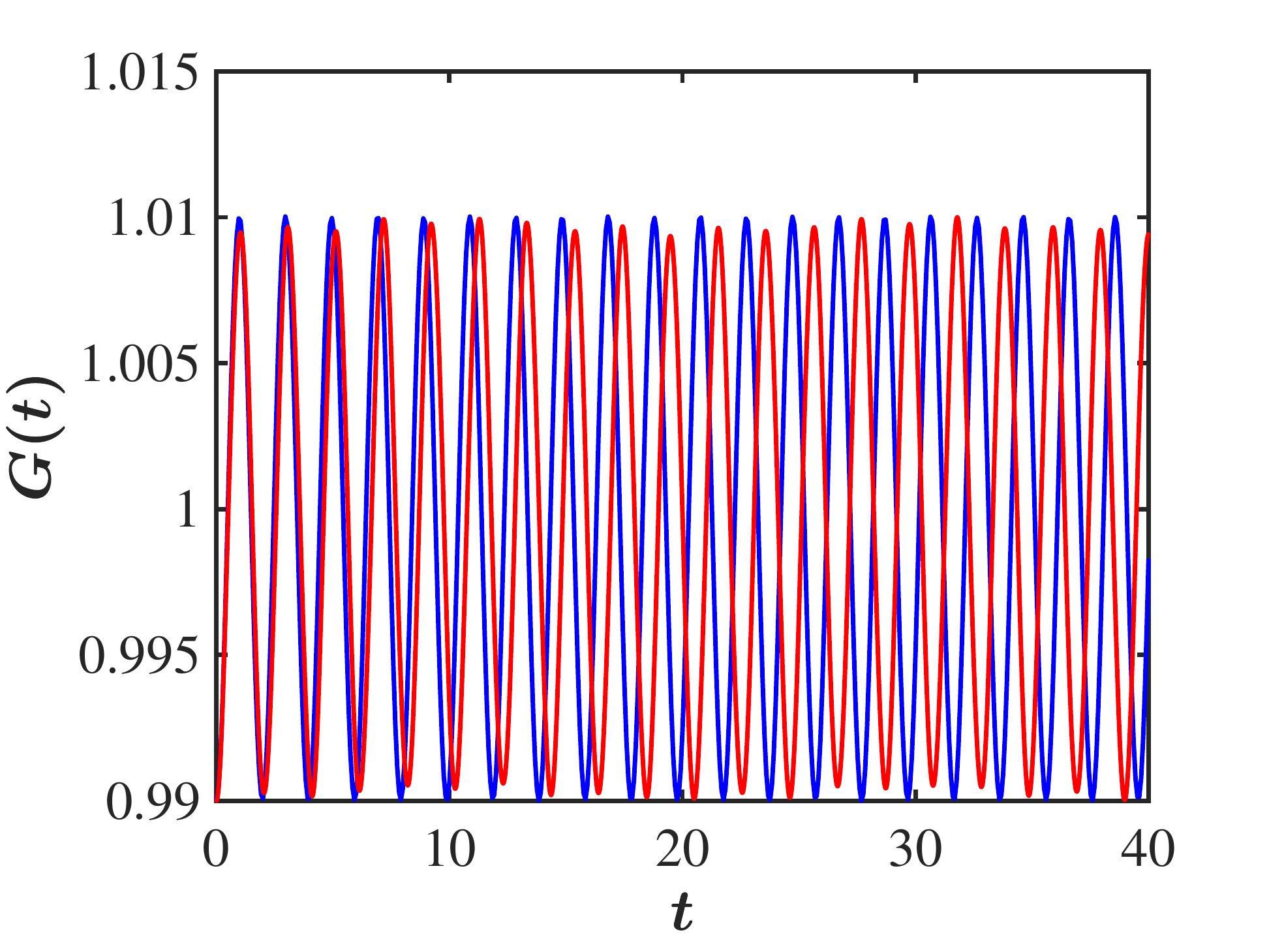}}
   \hspace{0.5em}
   \subfigure[$q(t)$ case b]
   {\label{f:q-Kappa-05-Case-b}
   \includegraphics[width=0.475\columnwidth]{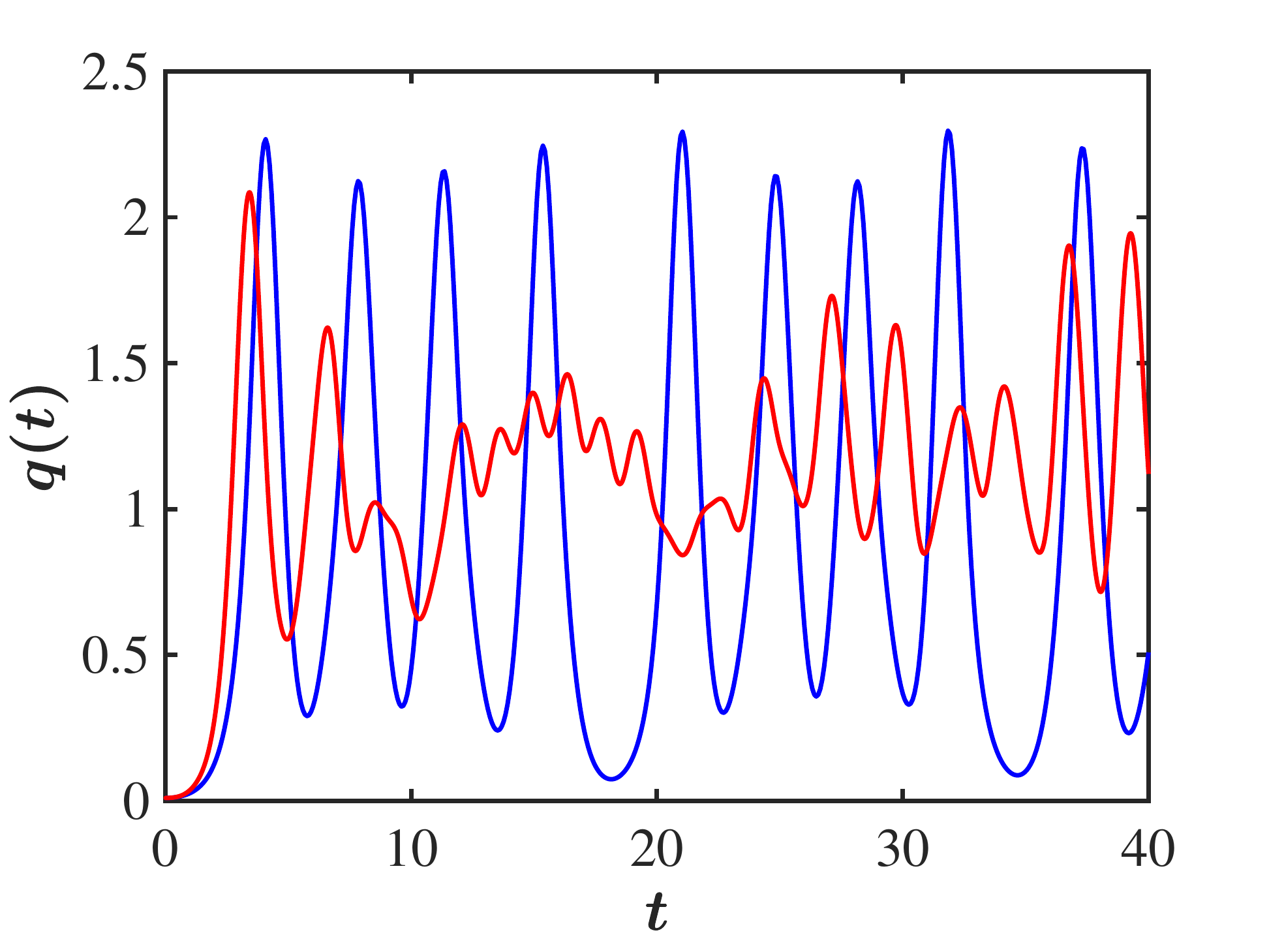}}
   \hspace{0.5em}
   \subfigure[$G(t)$ case b]
   {\label{f:G-Kappa-05-Case-b}
   \includegraphics[width=0.475\columnwidth]{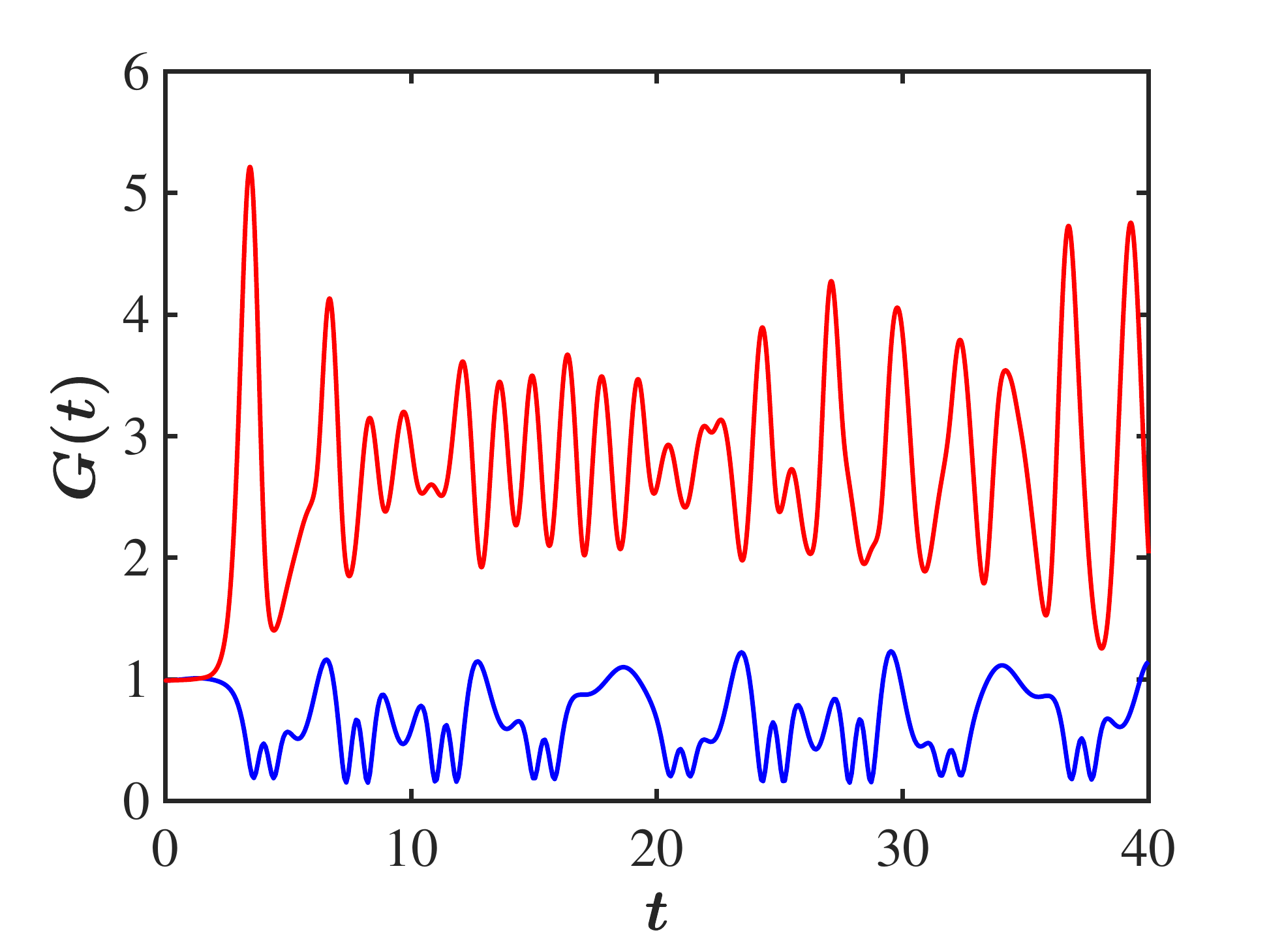}}
   \hspace{0.5em}
   \subfigure[$q(t)$ case c]
   {\label{f:q-Kappa-05-Case-c}
   \includegraphics[width=0.475\columnwidth]{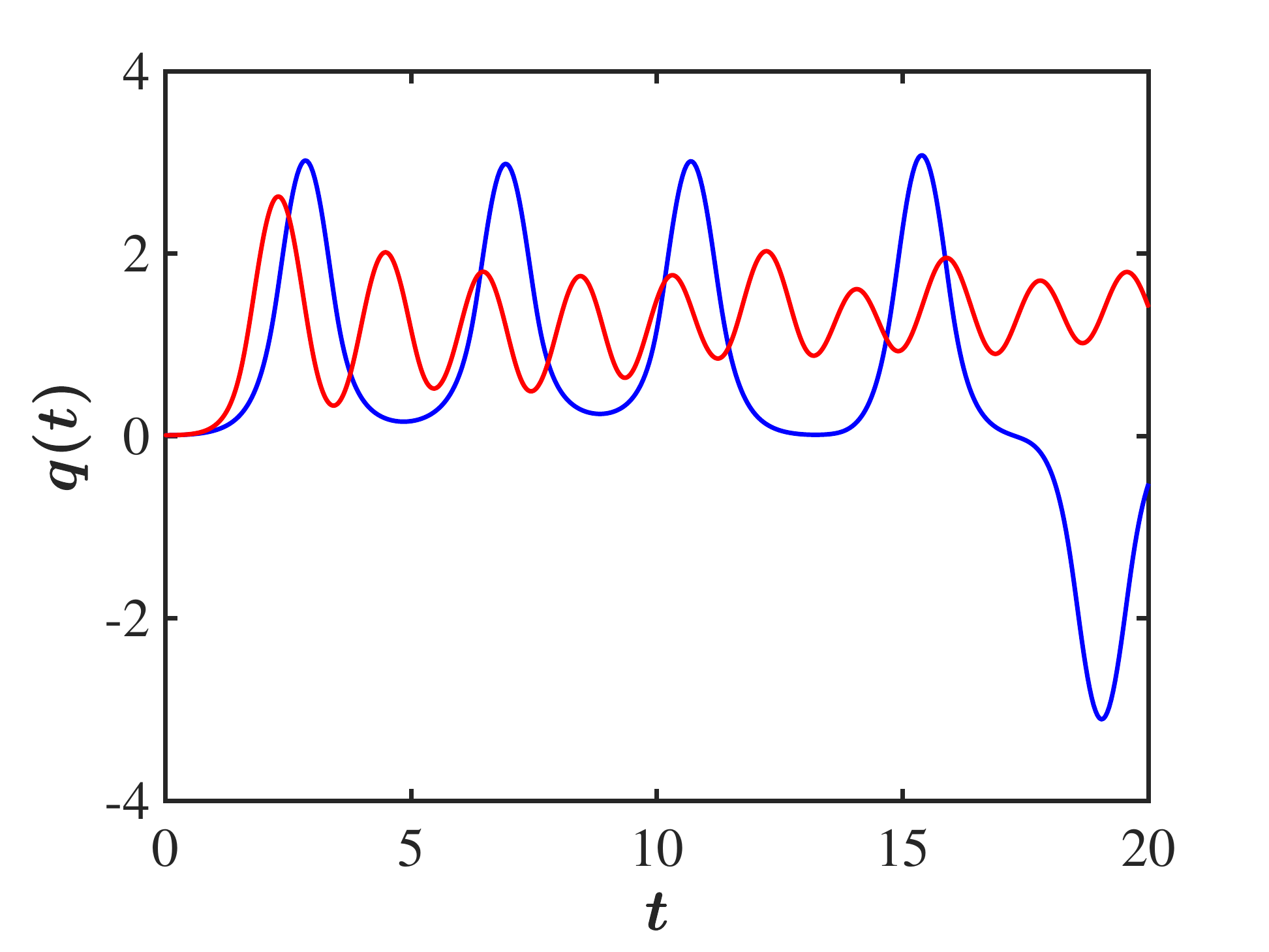}}
   \hspace{0.5em}
   \subfigure[$G(t)$ case c]
   {\label{f:G-Kappa-05-Case-c}
   \includegraphics[width=0.475\columnwidth]{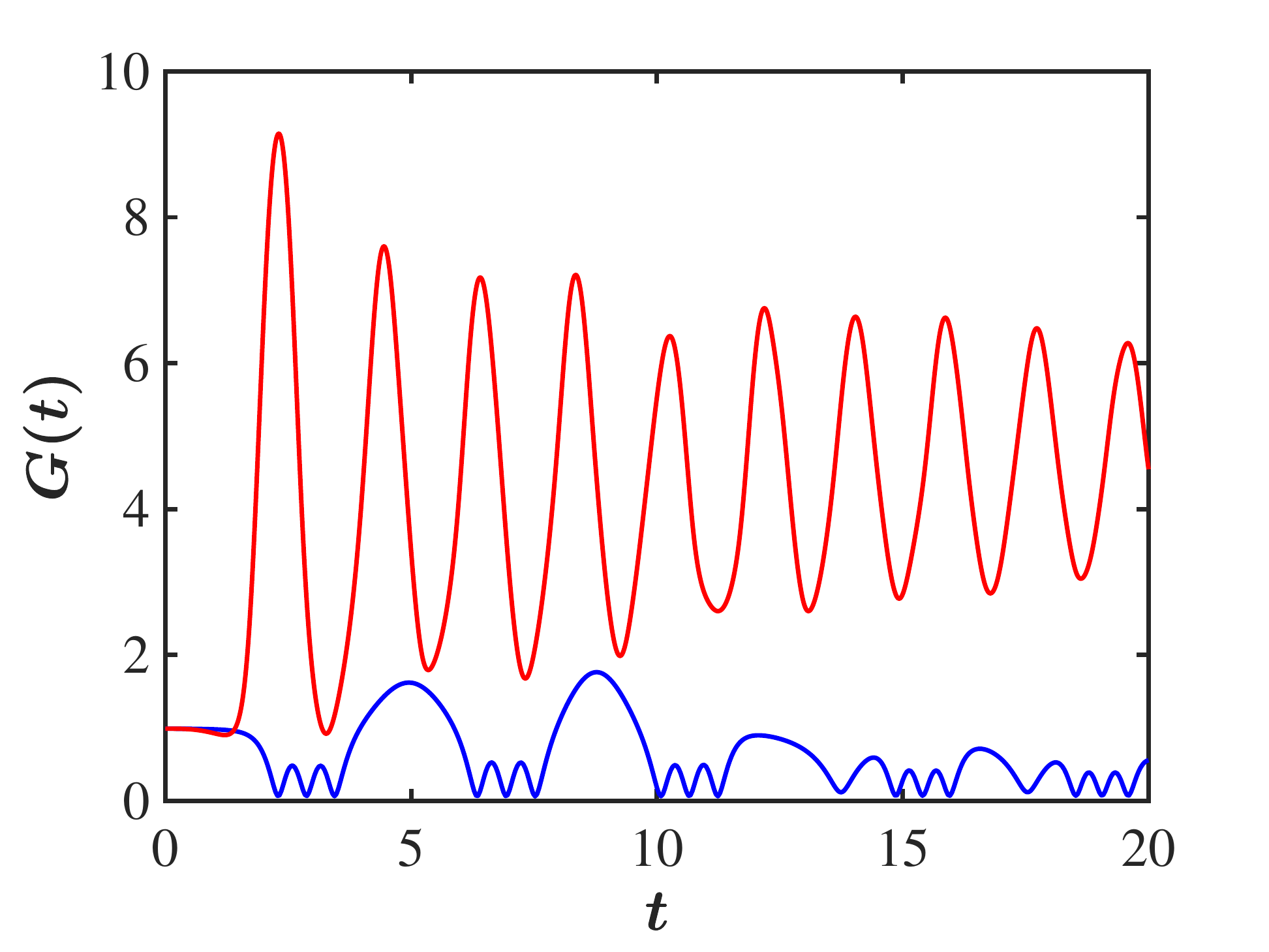}}
   \caption{\label{f:Kappa-05} Plots of $q(t)$ and $G(t)$ for the
   4CC results (blue) and numerical NLSE results (red),
   for $\kappa=1/2$.  See Table~\ref{t:plotdata} for mass parameters.}  
\end{figure}

Case (a): Using the data of Table~\ref{t:plotdata}, the solutions 
for $q(t)$ and $G(t)$ never go unstable in the absence of an external 
potential.  With an external potential we get the results shown in 
Fig.~\ref{f:q-Kappa-05-Case-a}.  The oscillation frequencies here 
match the prediction of the 4CC small amplitude approximation. 
%
%
%
%

\noindent
Case (b): Here $q(t)$  tries to escape the potential well, and $G$ 
tries to go to zero (blowup)  or infinity (collapse). However, energy 
conservation prevents both blowup and escape of the initial wavefunction, 
and we get the semi-oscillating behavior shown in Fig.~\ref{f:q-Kappa-05-Case-b}.


\noindent
Case (c): Similarly, 
blowup of $G(t)$ is stalled because of energy conservation.
The $q(t)$ growth also stalls, and $q(t)$ switches from being greater than zero 
to being less than zero.  This is seen in Fig.~\ref{f:q-Kappa-05-Case-c}.  

%
%
\subsection{$\kappa = 1$}

\begin{figure}[t]
   \centering
   \subfigure[$q(t)$ case a]
   {\label{f:q-Kappa-10-Case-a}
   \includegraphics[width=0.475\columnwidth]{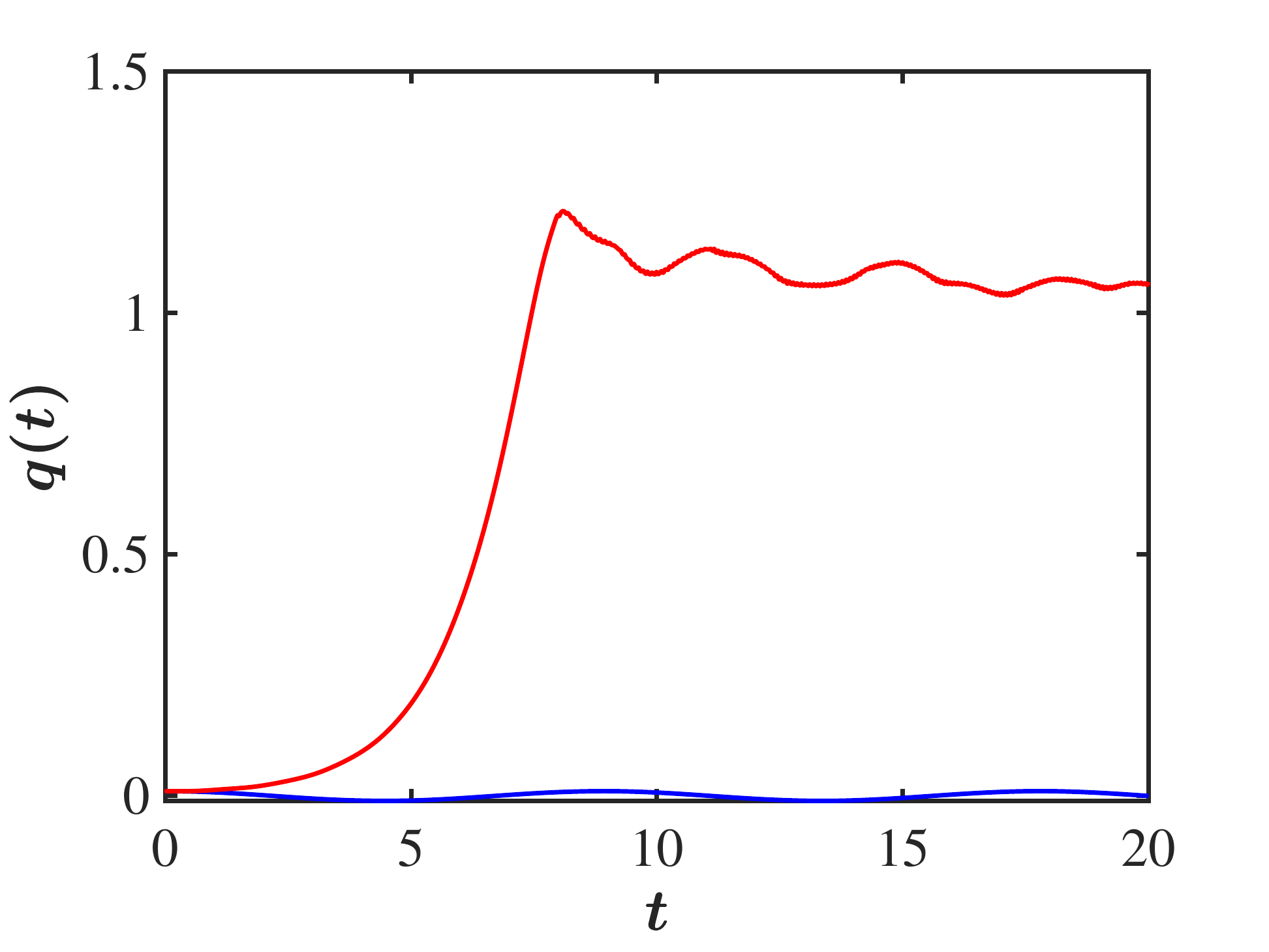}}
   \hspace{0.5em}
   \subfigure[$G(t)$ case a]
   {\label{f:G-Kappa-10-Case-a}
   \includegraphics[width=0.475\columnwidth]{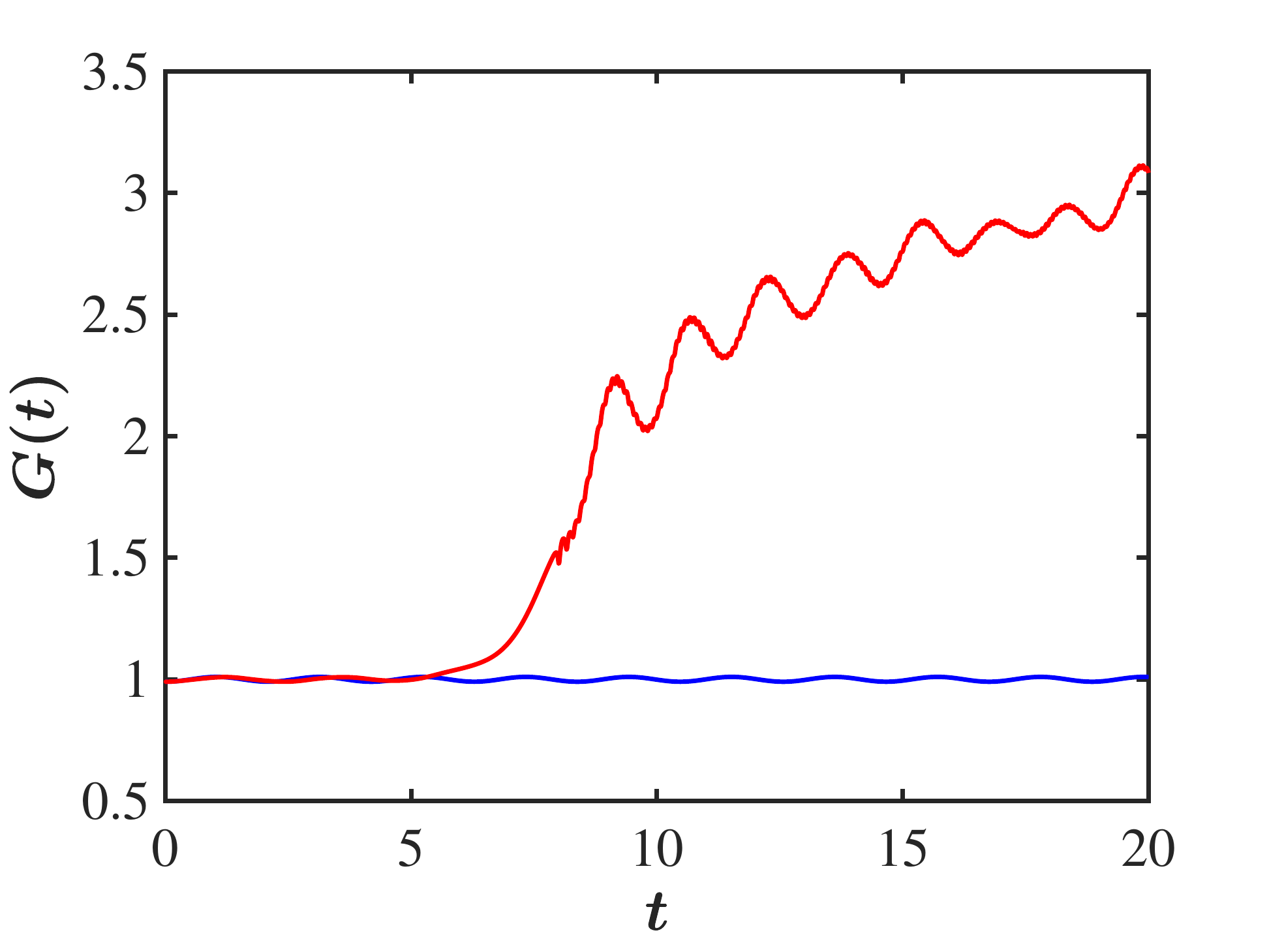}}
   \hspace{0.5em}
   \subfigure[$q(t)$ case b]
   {\label{f:q-Kappa-10-Case-b}
   \includegraphics[width=0.475\columnwidth]{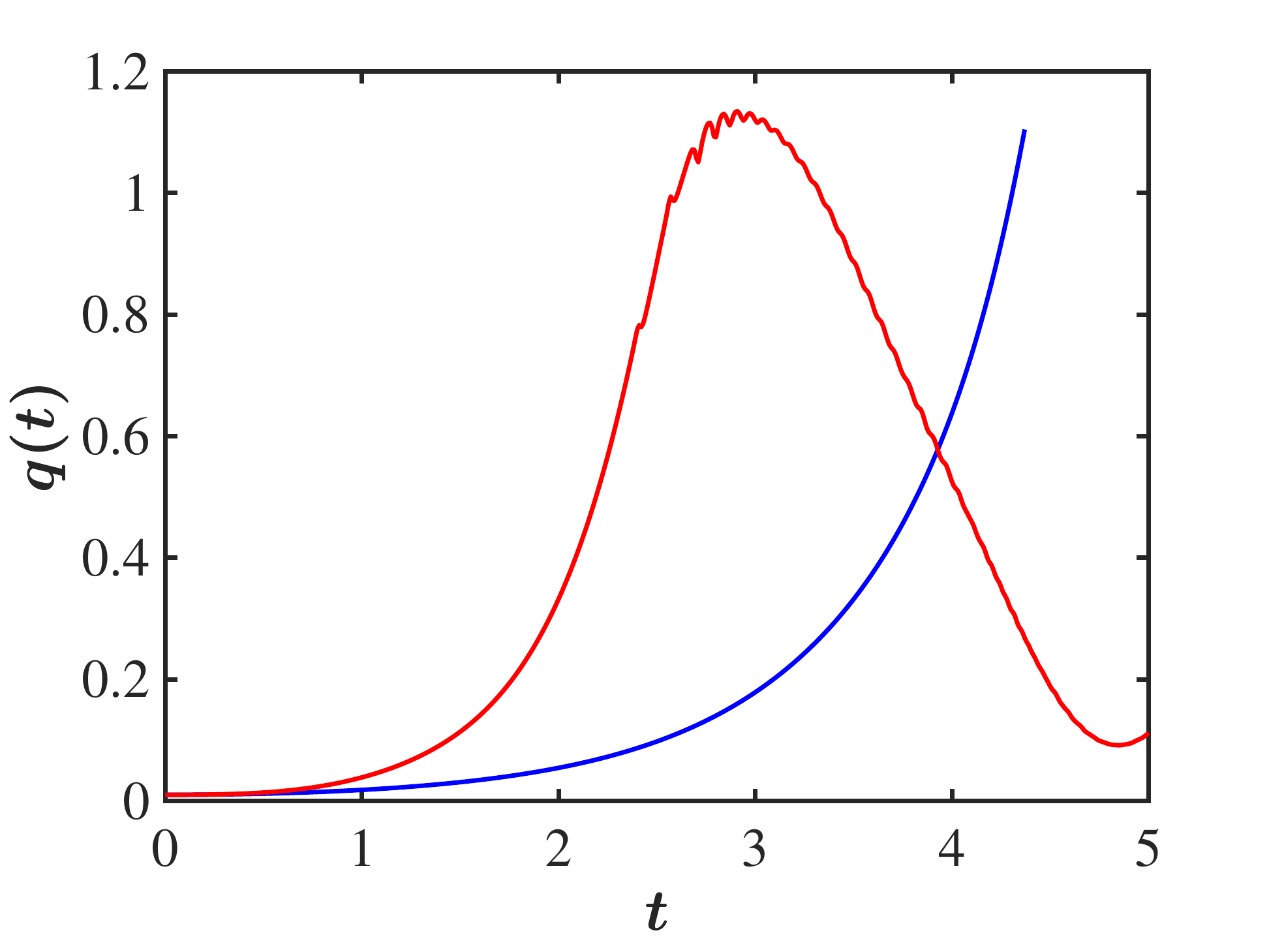}}
   \hspace{0.5em}
   \subfigure[$G(t)$ case b]
   {\label{f:G-Kappa-10-Case-b}
   \includegraphics[width=0.475\columnwidth]{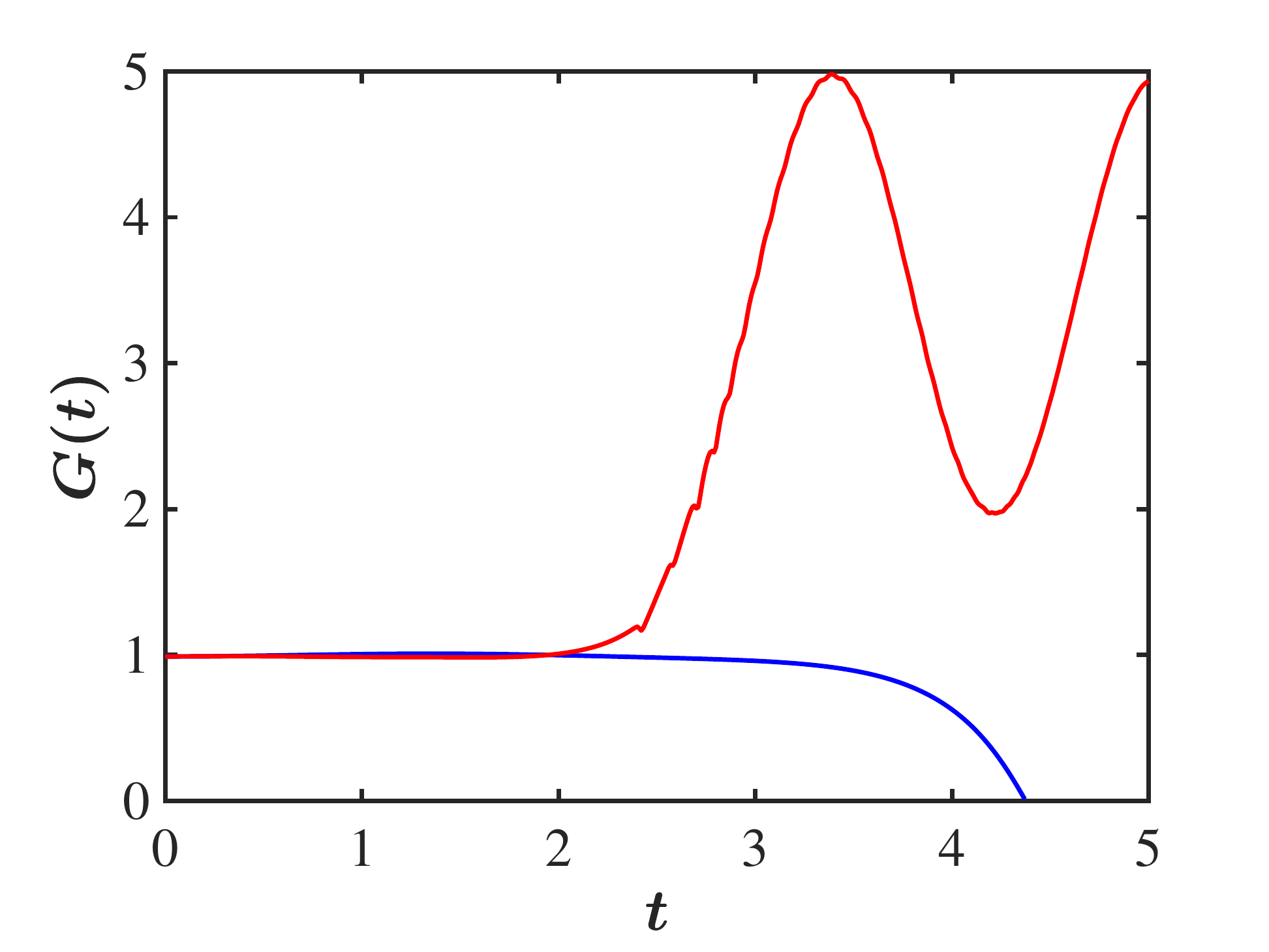}}
   \hspace{0.5em}
   \subfigure[$q(t)$ case c]
   {\label{f:q-Kappa-10-Case-c}
   \includegraphics[width=0.475\columnwidth]{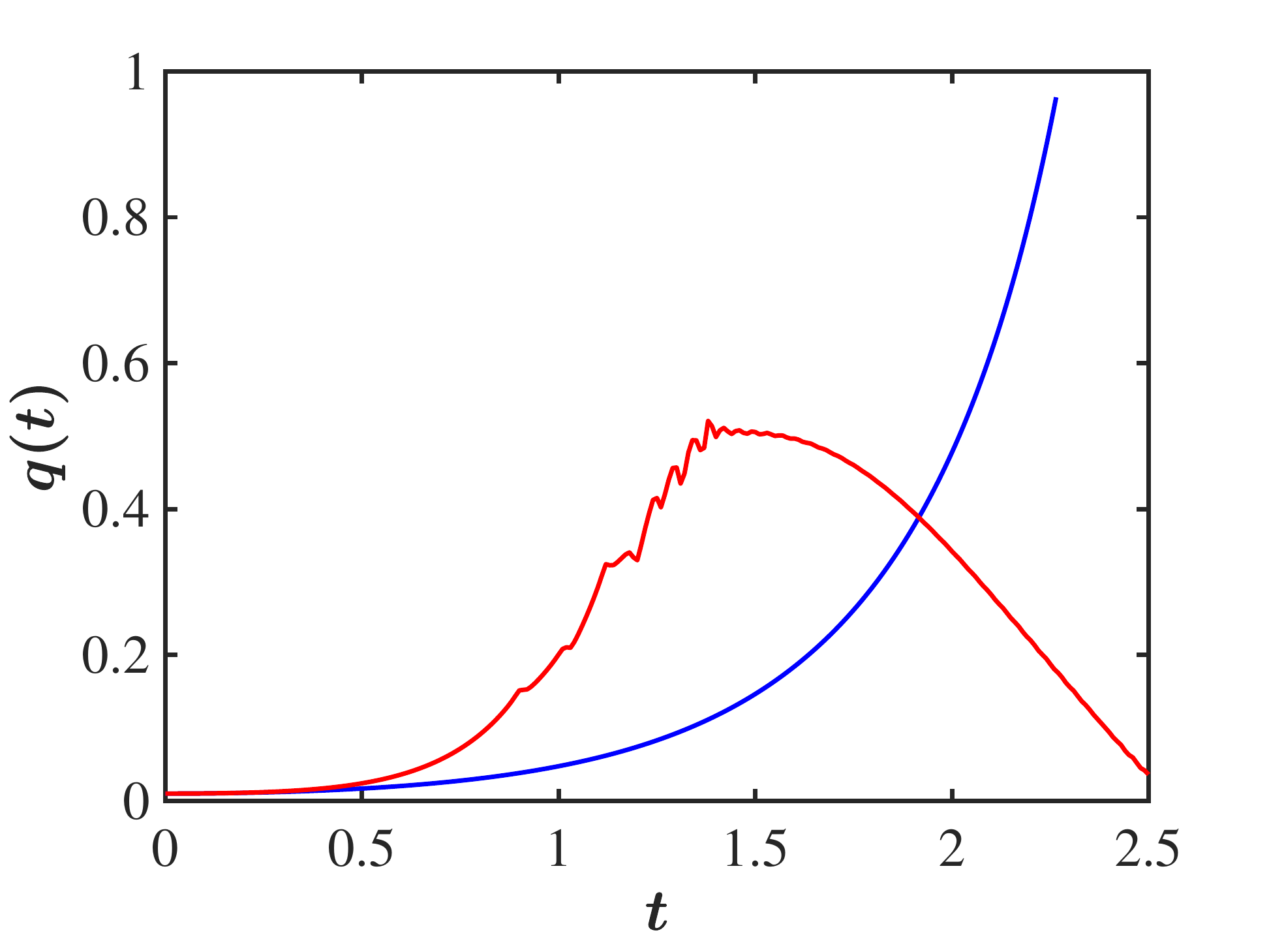}}
   \hspace{0.5em}
   \subfigure[$G(t)$ case c]
   {\label{f:G-Kappa-10-Case-c}
   \includegraphics[width=0.475\columnwidth]{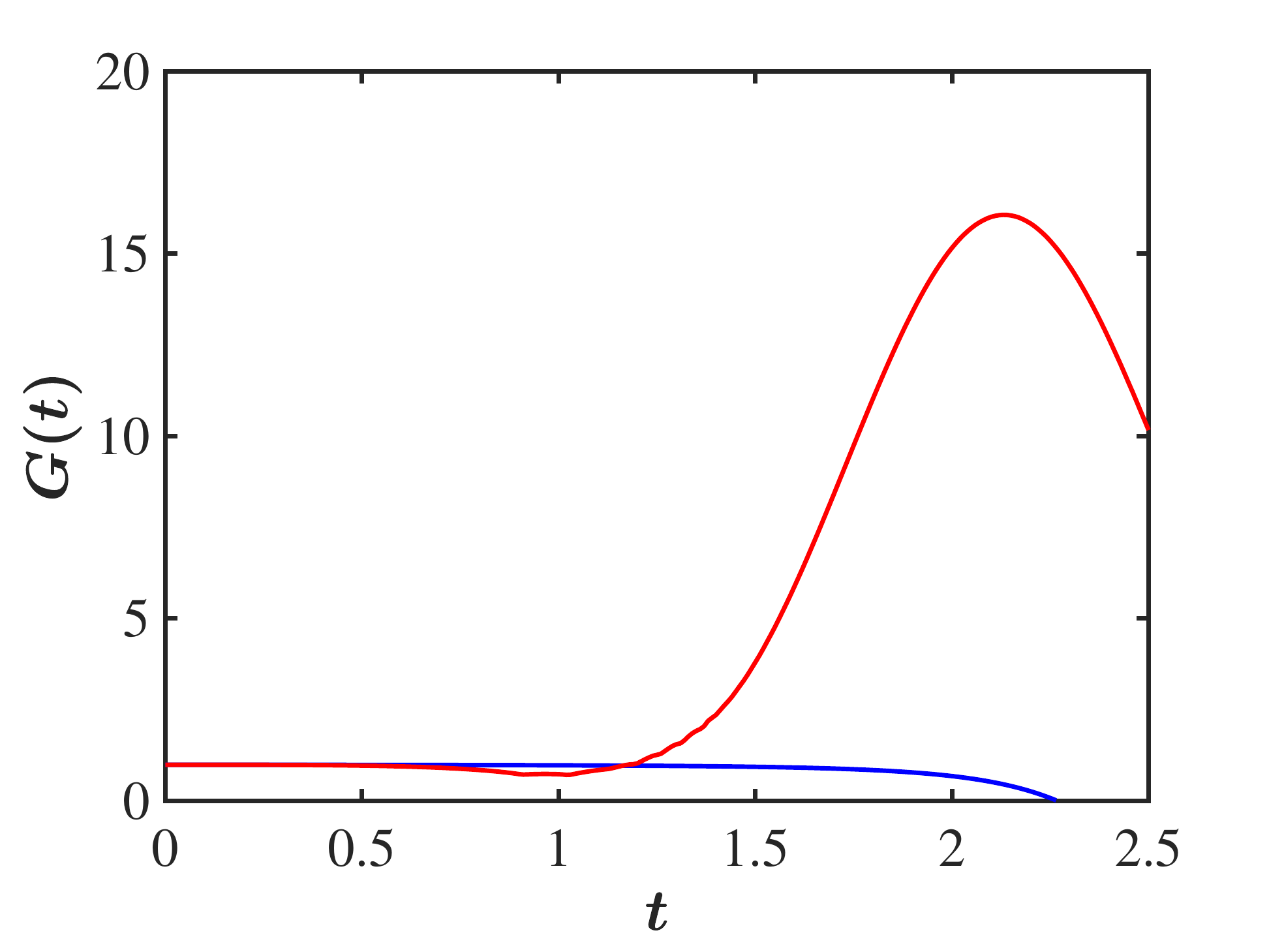}}
   \caption{\label{f:Kappa-10} Plots of $q(t)$ and $G(t)$ for the
   4CC results (blue) and numerical NLSE results (red),
   for $\kappa=1$.  See Table~\ref{t:plotdata} for mass parameters.}  
\end{figure}

Case (a): In this case, $\kappa=1$ is the critical value for blowup 
in the absence of a confining potential. Moreover, 
blowup occurs in this case when for the initial conditions $M \geq  M_{\mathrm{w}}$ holds. 
The 4CC results of Fig.~\ref{f:q-Kappa-10-Case-a} show that $q(t)$ and $G(t)$ 
oscillate, and are in the small amplitude regime. The period for $G(t)$ 
from the small amplitude approximation is $T_G = 2.118$ and the period 
for $q(t)$ is $T_q = 9.934$. 
%
%
%
%
\noindent
Case (b): If we are in the in-between case, then after one oscillation
of the $G(t)$ variable, the wavefunction blows up as a result of the 
$q(t)$ instability.  This is seen in Fig.~\ref{f:q-Kappa-10-Case-b} for 
$G(t)$ and $q(t)$.   


\noindent
Case (c): When we are above the critical mass, the solution blows up much quicker. 
For this case, the blowup time is shortened to about $t_f=2.3$, which is seen in 
Fig.~\ref{f:q-Kappa-10-Case-c}.   
%
%
\subsection{$\kappa = 3/2$}

\begin{figure}[t]
   \centering
   \subfigure[$q(t)$ case a]
   {\label{f:q-Kappa-15-Case-a}
   \includegraphics[width=0.475\columnwidth]{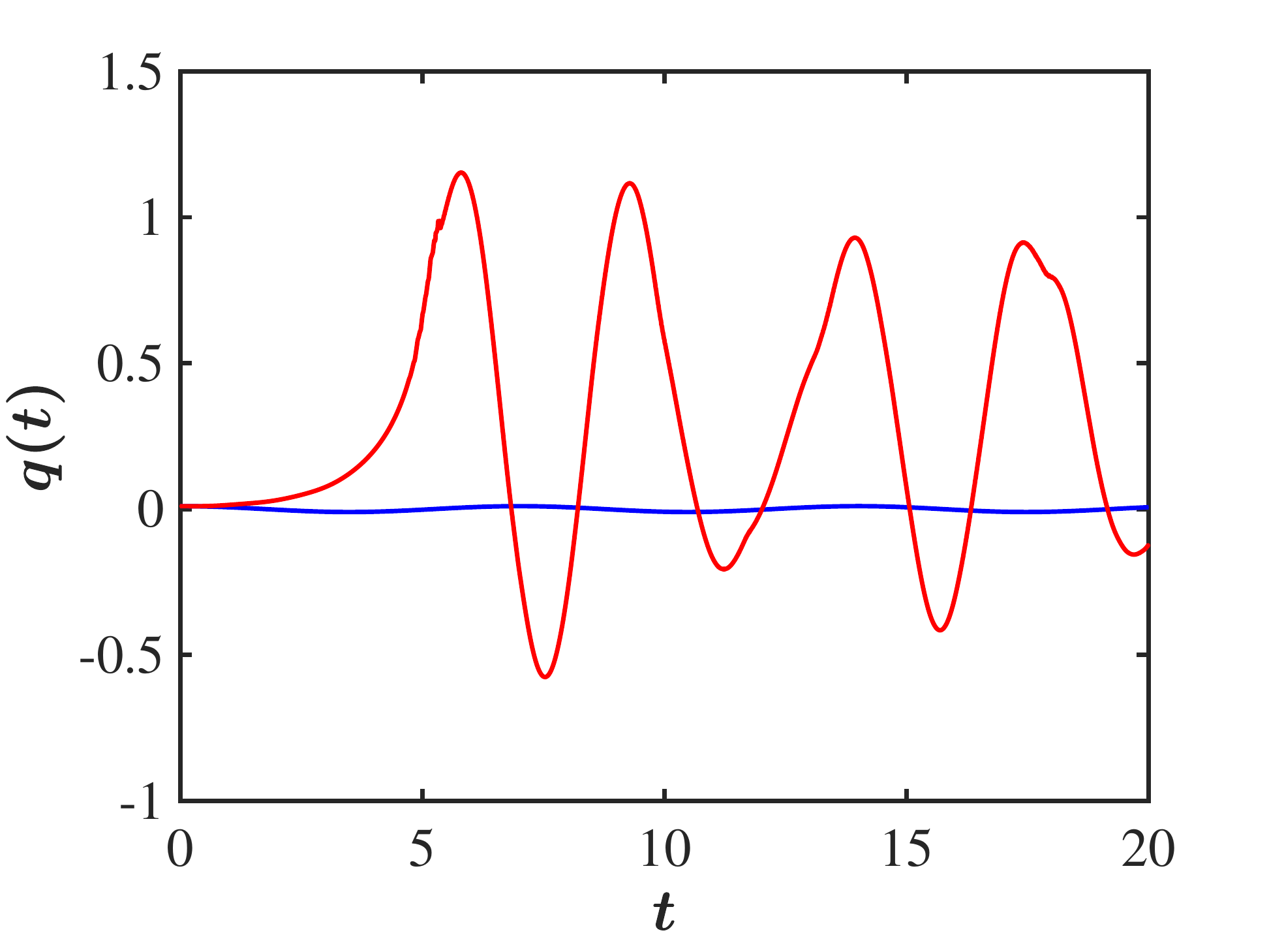}}
   \hspace{0.5em}
   \subfigure[$G(t)$ case a]
   {\label{f:G-Kappa-15-Case-a}
   \includegraphics[width=0.475\columnwidth]{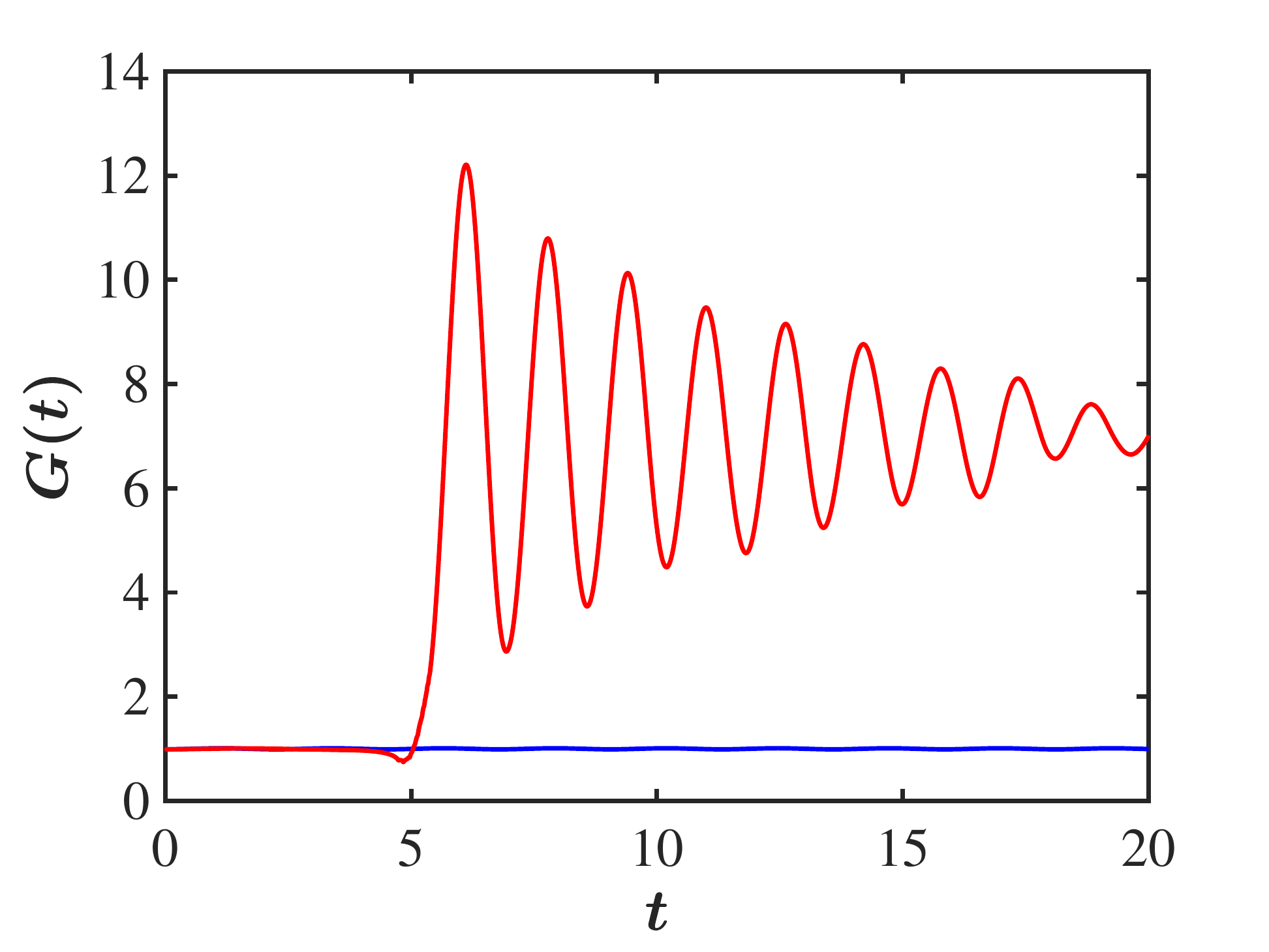}}
   \hspace{0.5em}
   \subfigure[$q(t)$ case b]
   {\label{f:q-Kappa-15-Case-b}
   \includegraphics[width=0.475\columnwidth]{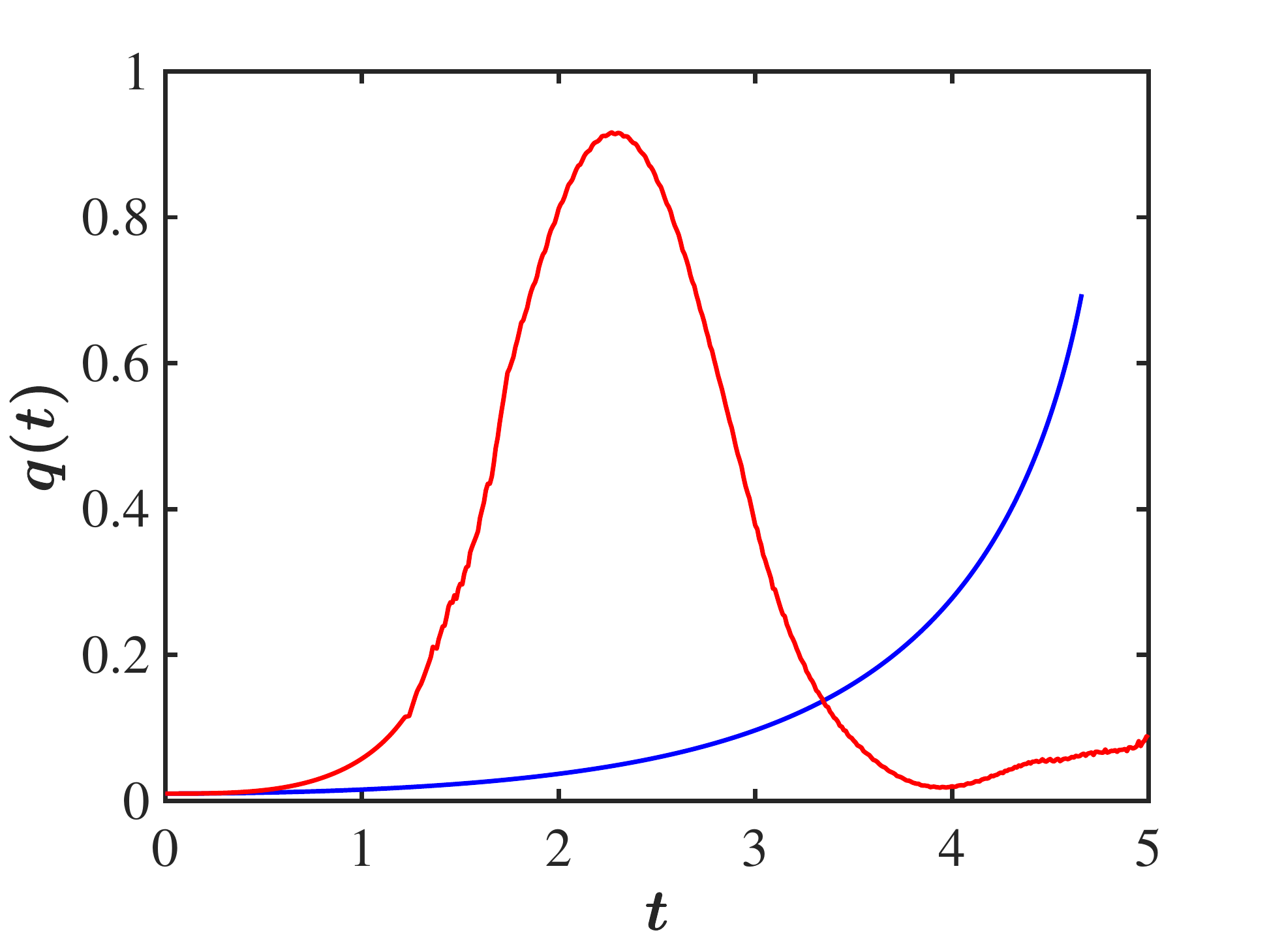}}
   \hspace{0.5em}
   \subfigure[$G(t)$ case b]
   {\label{f:G-Kappa-15-Case-b}
   \includegraphics[width=0.475\columnwidth]{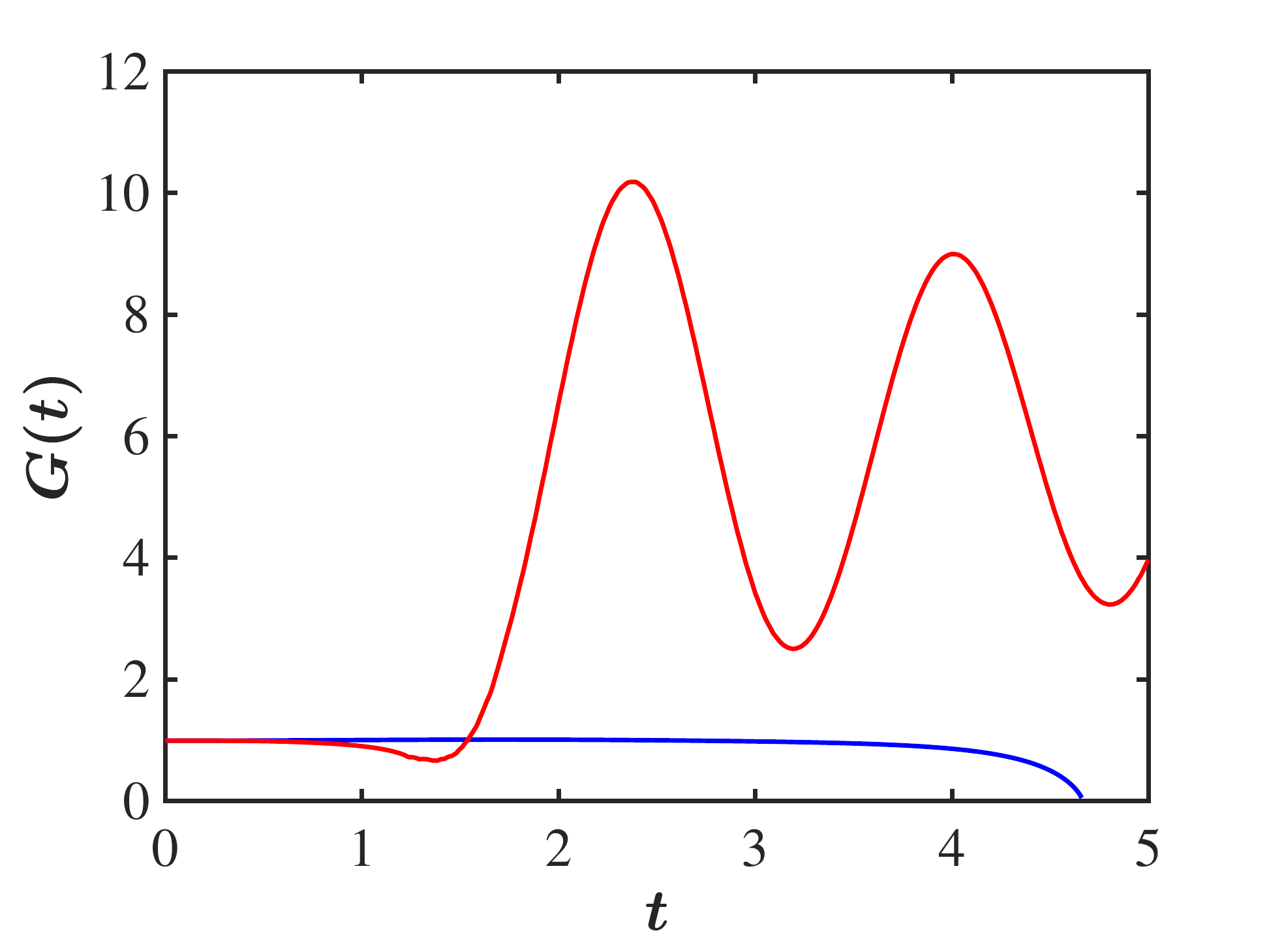}}
   \hspace{0.5em}
   \subfigure[$q(t)$ case c]
   {\label{f:q-Kappa-15-Case-c}
   \includegraphics[width=0.475\columnwidth]{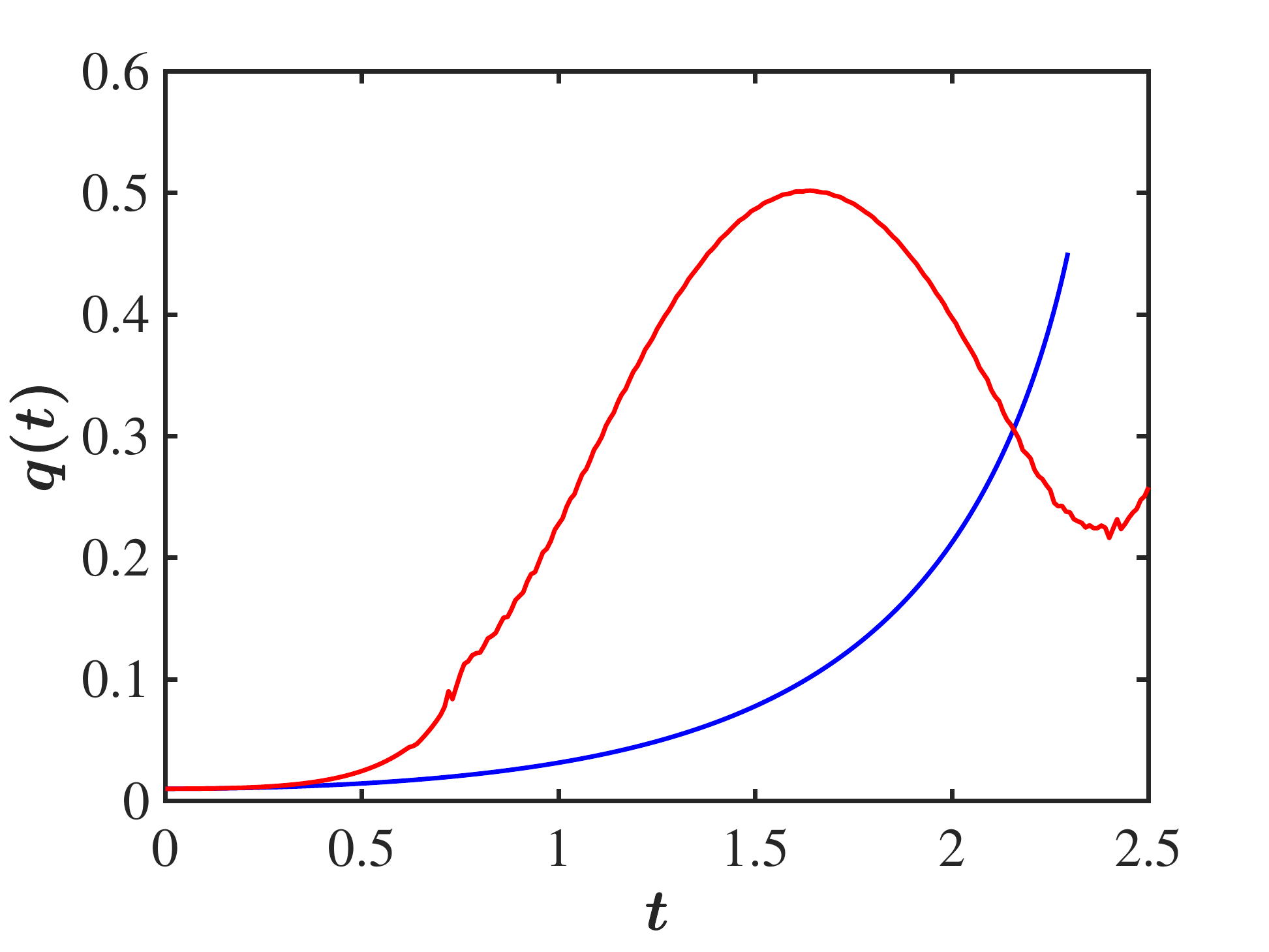}}
   \hspace{0.5em}
   \subfigure[$G(t)$ case c]
   {\label{f:G-Kappa-15-Case-c}
   \includegraphics[width=0.475\columnwidth]{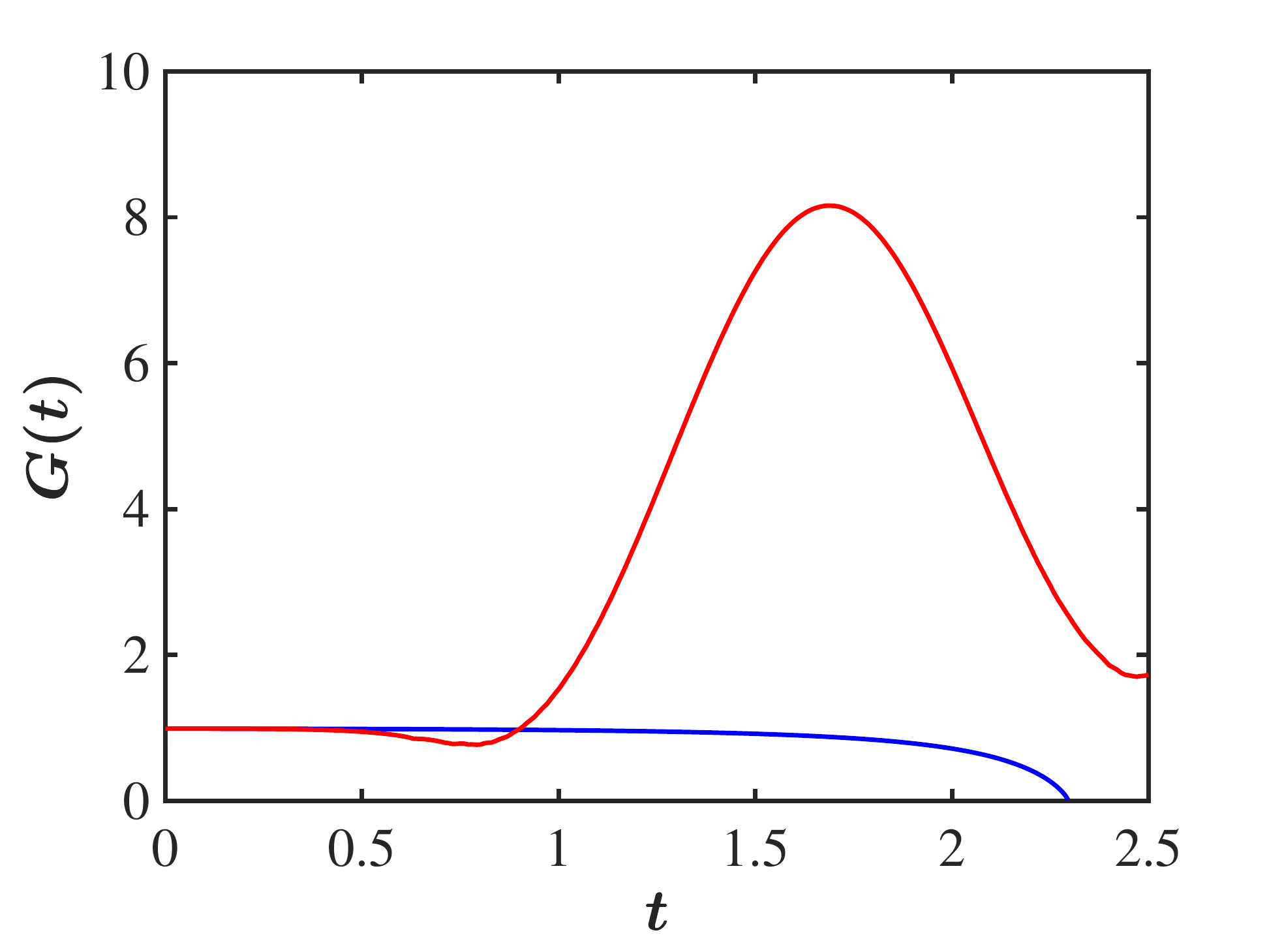}}
   \caption{\label{f:Kappa-15} Plots of $q(t)$ and $G(t)$ for the
   4CC results (blue) and numerical NLSE results (red),
   for $\kappa=3/2$.  See Table~\ref{t:plotdata} for mass parameters.}  
\end{figure}

\noindent
Case (a): In the absence of a confining potential, when $\kappa=3/2$ we 
are always in a blowup regime. However \emph{with} a confining potential, 
the 4CC results shown in Fig.~\ref{f:q-Kappa-15-Case-a} indicate that we 
are in a small amplitude regime. The two periods are predicted from the 
small amplitude approximation are $T_q = 8.21$ and $T_G = 2.35$, which 
agree quite well with simulations. 
%
%
%
%
\noindent
Case (b): The results of the 4CC simulation for $G(t)$ and $q(t)$ are 
shown in Fig.~\ref{f:q-Kappa-15-Case-b}, where we find that $q(t)$ is 
unstable but $G(t)$ is initially stable for one period and then the 
wavefunction blows up at $t \approx 4$ as a result of the translation 
instability.

\noindent
Case (c): For this case, we see from Figs.~\ref{f:q-Kappa-15-Case-c} 
and~\ref{f:G-Kappa-15-Case-c} that both $q(t)$ and $G(t)$ blow up quicker,
and the blowup happens at $t \approx 3.5$.

%
%
\begin{figure}[t]
   \centering
   \subfigure[$q(t)$ and $M=4.0$]
   {\label{f:q-Kappa-20-M40}
   \includegraphics[width=0.475\columnwidth]{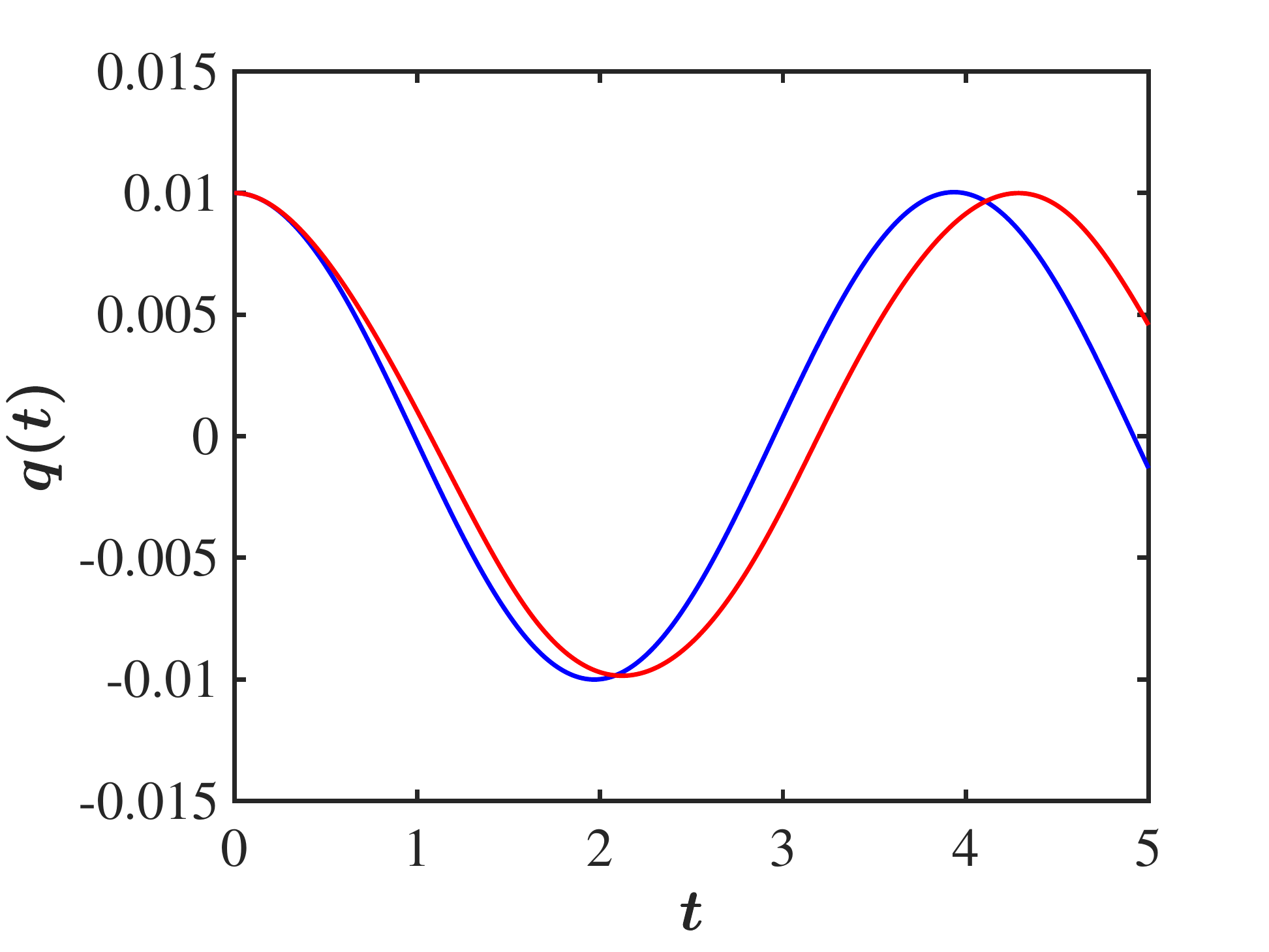}}
   \hspace{0.5em}
   \subfigure[$G(t)$ and $M=4.0$]
   {\label{f:G-Kappa-20-M40}
   \includegraphics[width=0.475\columnwidth]{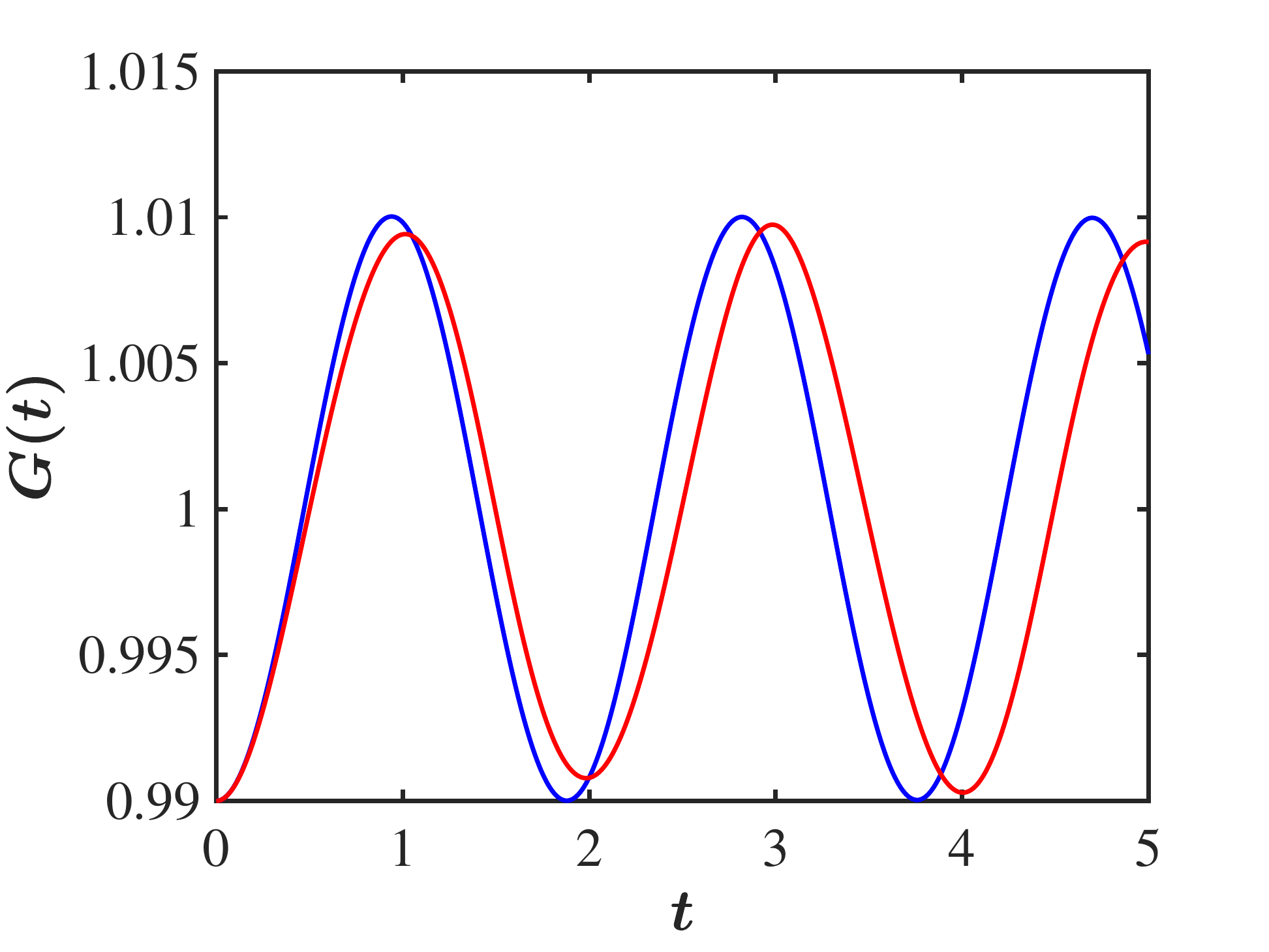}}
   \caption{\label{f:Kappa-20} Plots of $q(t)$ and $G(t)$ for the
   4CC results (blue) and numerical NLSE results (red),
   for $\kappa=2$. Here $M = 4.0$ with $A_0 = 1.341$. }  
\end{figure}
%
%

%
%

\subsection{$\kappa = 2$, stable regime}

In Fig.~\ref{f:Kappa-20} we show the results for $q(t)$ and $G(t)$ in 
the stability region where $\kappa = 2$ and $M = 4$.  The two methods 
give very similar results in this stable oscillatory  region.


%
%
\section{\label{s:stability}Numerical Stability Analysis}

We now turn our focus on the spectral stability analysis of stationary
solutions to the NLSE of Eq.~\eqref{NLSE-2D}. In doing so, we consider
first the separation of variables ansatz
\begin{equation}
   \psi(\mathbf{r},t) 
   = 
   \phi(\mathbf{r}) \, \rme^{-\rmi\omega t}
   \qc 
   \omega = 2/G_{0} \>,
\label{sep_ansatz}
\end{equation}
with $G_{0}=1$, and upon substituting Eq.~\eqref{sep_ansatz} into Eq.~\eqref{NLSE-2D},
we arrive at the steady-state problem:
\begin{equation}
   -
   \nabla^{2} \phi
   -
   g \, |\phi|^{2\kappa}\phi 
   +
   [\, V(\mathbf{r})-\omega \,] \, \phi
   =
   0
\label{steady_state_bvp}
\end{equation}
supplemented with zero Dirichlet boundary conditions (BCs), i.e., $\phi=0$ at 
infinity. It should be noted that the physical domain $\mathbb{R}^{2}$ is truncated 
into a finite one, i.e., $\Omega=[-L,L]^{2}$ with $L=15$ at which the zero Dirichlet 
BCs are imposed on $\partial\Omega$. Then, the computational domain $\Omega$ is discretized
homogeneously (i.e., with $\Delta x=\Delta y$) using $N=301$ points along each direction,
and the Laplacian appearing in Eq.~\eqref{steady_state_bvp} is replaced by a fourth-order
accurate, centered finite difference scheme. The resulting (large) system of nonlinear 
equations emanating from the above discretization method is solved by means of Newton's
method with tolerances (on both the iterates and nonlinear residual) of $10^{-13}$. The 
initial seed for Newton's method is provided by the exact solution of Eq.~\eqref{assume1}
for given $\kappa$, $g$, and $A$. Although the exact solution is available in our setup, 
we compute the numerically exact solution on the computational grid we employ since the 
former does not satisfy \textit{exactly} the discrete equations we obtain per the discretization 
scheme considered herein due to local truncation error. 

Having identified a steady-state solution, we perform a two-parameter continuation on
the $(\kappa,A)$-plane, and compute branches of solutions. We perform a spectral stability
analysis, i.e., Bogoliubov de-Gennes (BdG) analysis \cite{deGennes-1966}, of the pertinent 
states at each continuation step by considering the perturbation ansatz around a steady-state 
$\phi^{(0)}(\mathbf{r})$ of the form
\begin{equation}\label{pertr_ansatz}
   \tilde{\psi}(\mathbf{r},t)
   =
   \rme^{-\rmi\omega t} \,
   \left[ \,
      \phi^{(0)}
      +
      \varepsilon 
      \left ( \,
         a(\mathbf{r}) \, \rme^{\lambda t}
         +
         b^{\ast}(\mathbf{r}) \, \rme^{\lambda^{\ast}t} \,
      \right ) \,
   \right] 
   \qc
   \varepsilon \ll 1 \>.
\end{equation}
Upon plugging Eq.~\eqref{pertr_ansatz} into Eq.~\eqref{NLSE-2D}, we arrive 
(at order ${\mathcal{O}}(\varepsilon)$) at the eigenvalue problem:
\begin{equation}
   \Bigl ( \begin{array}{cc}
      A_{11}         &  A_{12} \\
      -A_{12}^{\ast} & -A_{11}^{\ast}
   \end{array} \Bigr )
  \Bigl ( \begin{array}{c}
      a\\
      b
   \end{array} \Bigr )
   =
   \rmi \, \lambda \,
   \Bigl ( \begin{array}{c}
      a\\
      b
   \end{array} \Bigr ) \>,
\end{equation}
whose matrix elements are given by:
\begin{eqnarray}
   A_{11}
   &=&
   -\nabla^{2}-g\left(\kappa+1\right)|\phi^{(0)}|^{2\kappa}+V-\omega,
   \\
   A_{12}
   &=&
   -g\kappa|\phi^{(0)}|^{2\kappa-2}\left(\phi^{(0)}\right)^{2}.
\end{eqnarray}
A solution is deemed linearly stable if all the eigenvalues $\lambda=\lambda_{r}+\rmi \lambda_{i}$ 
lie on the imaginary axis (i.e., $\lambda_{r}\equiv 0$). On the other hand, if an eigenvalue 
$\lambda$ has a non-zero real part, that signals an instability and thus the solution is deemed
(linearly) unstable.

\begin{figure}
   \centering
   \includegraphics[width=0.6\linewidth]{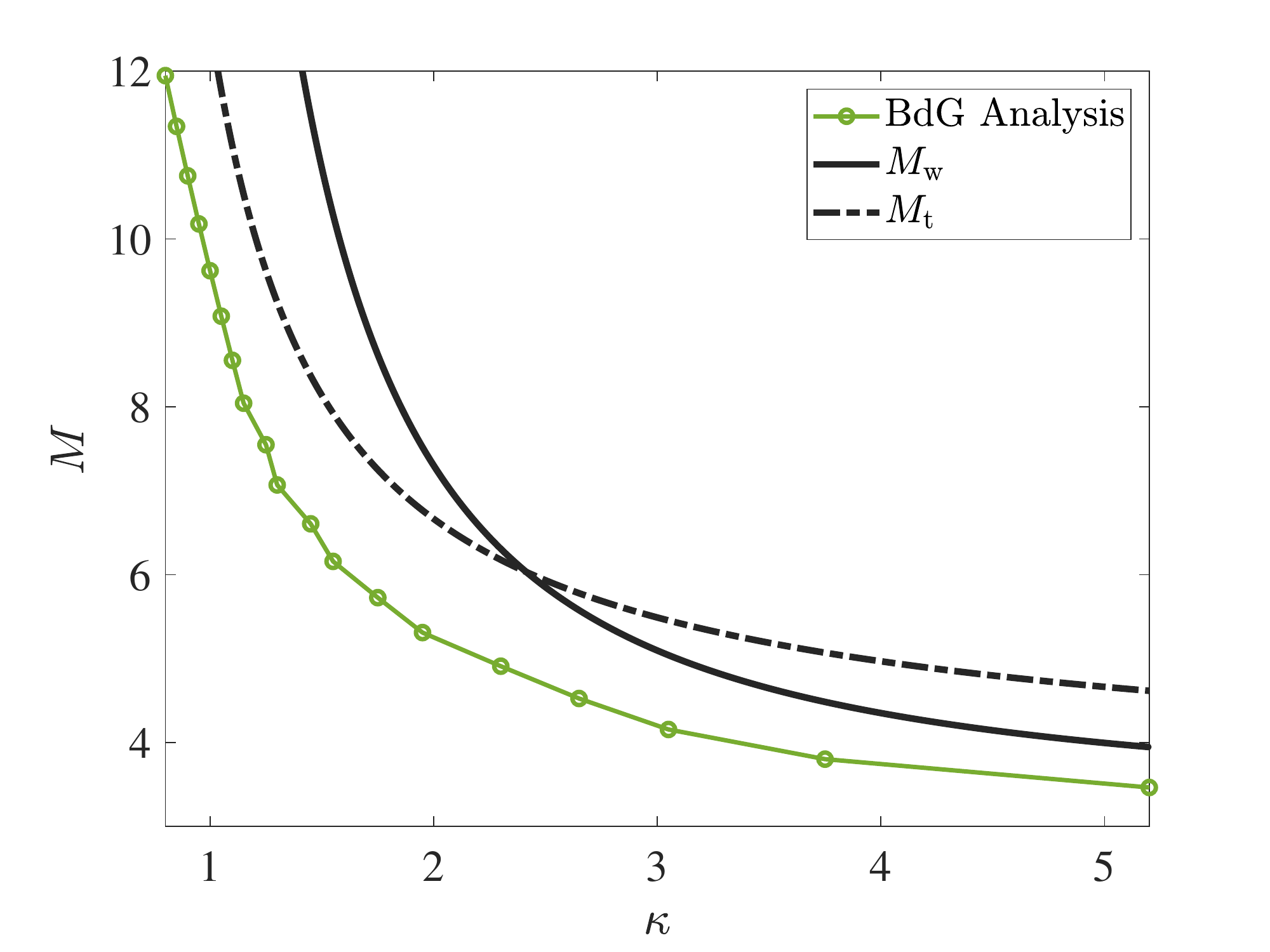}
   \caption{$M^{\ast}$ vs. $\kappa$. The two curves from Derrick's Theorem lie above the numerically 
   determined curve for instability to set in. See also, Fig.~\ref{fig:Mcrit-a}.} 
\label{fig:Mstar3}
\end{figure}

We have performed a systematic spectral stability analysis on the $(\kappa,A)$-plane whence at 
the points at which the solution is spectrally unstable, we calculated the total mass given by 
Eq.~\eqref{mass}. Our numerical results ($M$ vs $\kappa$) are shown with the green curve in 
Fig.~\ref{fig:Mstar3} where we also graphed the two critical mass curves for comparison (see the 
legend therein). What we find is that the onset of instability lies on a curve {\em below} the two 
critical mass curves found by Derrick's theorem.  This is quite different from the result found 
for the $(1+1)$-dimensional NLSE in a P{\"o}schl-Teller external potential \cite{Dawson-2017} where 
the numerical curve lies above the curve found by Derrick's theorem. 
%
%
%
%

\section{Conclusions}
\label{s:conclusions}
%
%
  
In this paper we have revisited the problem of blowup in the nonlinear 
\Schrodinger\ equation with arbitrary nonlinearity exponent $\kappa$.
In particular, we used the result that an arbitrary initial ground state 
wavefunction can be converted into an exact solution if we place it in 
a well-determined external potential. We find that in this confining potential 
the wavefunction can become unstable to both width and translation perturbations. 
There are two different onset masses at which this happens, with the translational 
instability occurring at a \emph{lower/higher} mass than the width instability 
depending on whether $\kappa$ is less than or greater than $ 1+\sqrt{2}$.  
The numerical BdG analysis gives a curve for the critical mass that lies 
slightly below both these curves although it follows a similar trend. In the 
4CC variational approximation, there are now several  regimes with quite 
different behavior. When there is no confining potential, $\kappa d \geq 2$ 
defines the regime where there can be a blowup. 

The case we study in detail here is $d=2$, so that $\kappa=1$ is the critical 
value of $\kappa$.  What we find for the case when we are in the confining 
potential, when $\kappa < 1$, one can not have blowup (or collapse)  because 
of energy conservation, but there are now three distinct regimes.  When we are 
below both critical masses, there is a regime of small oscillation response to 
small perturbations of the initial conditions. As we cross the threshold for $q$ 
instability, then we can have ``frustrated'' blowup where $G$ first oscillates and 
then the growth of $q$ causes the wavefunction to start spiking. However energy 
conservation prevents blowup from completing. Then one gets a sort of repetition
of this pattern.  When one crosses the second instability, there is a combination of 
oscillatory regions at low $G$ combined with peaking and collapsing. 

The wavefunction can also oscillate about different values of $q$ both positive and 
negative. Once $\kappa \geq 1$, we have mainly two regimes. When we are below the 
two critical masses, we have oscillatory response to small perturbations. Once the 
$q$ instability is present, it then triggers blowup of the wavefunction. When 
one is below the second critical mass, the width makes one oscillation before one 
starts the blowup regime, as $q$ increases exponentially in time. We expect these types 
of behavior to exist irrespective of the exact choice of the initial approximate wavefunction 
used to describe the soliton in the absence of the external potential.  In the stable 
regime, which is the small oscillation regime of the variational approximation, agreement 
with numerical simulation of the NLSE is quite good. However in the unstable regime, once 
the values of the first and second moments of the wavefunction start deviating in a substantial 
way from their initial values, other degrees of freedom get excited and our simple 4CC ansatz 
does not capture the behavior of the wavefunction very well.


%
%
\appendix
%
%
\section{\label{s:extend} Extension to arbitrary dimension}
%
%
%

In an arbitrary number of spatial dimensions $d$, one can assume 
arbitrary finite norm initial data and again find the potential 
that will lead to this initial data being an exact solution. If 
we take the initial data to be of the form
\begin{equation}
   \psi(r,0) = A \, u(r)\>,
\end{equation}
where $A$ is the amplitude, and then assume that the time-dependent 
solution is of the form:
\begin{equation}
   \psi(r,t) = A \, u(r) \, \rme^{- \rmi \, \omega t} \>.
\end{equation}
Then since the Laplacian in $d$ dimensions for radial solutions is
\begin{equation}
   \nabla^{2} \psi(r,t)
   =
   \pddv{\psi(r,t)}{r} + \frac{(d-1)}{r} \pdv{\psi(r,t)}{r} \>,
\end{equation}
we find from \eqref{NLSE-2D} that $u(r)$ satisfies
\begin{equation}\label{A:NLSE}
   \omega + \frac{u''}{u} + \frac{(d-1)}{r} \frac{u'}{u} + g \, A^{2 \kappa} \, u^{2 \kappa} = V(r) \>.
\end{equation}
By choosing
\begin{equation}\label{A:omega} 
  \omega 
  + 
  \Bigl [\,
     \frac{u''}{u} + \frac{(d-1)}{r} \frac{u'}{u} \,
  \Bigr ]_{r=0}
  =
  0 \>, 
\end{equation}
we are able to remove the constant term from the potential when $g=0$.  
This way, upon solving Eq.~\eqref{A:omega} for $\omega$ and substituting 
this back into Eq.~\eqref{A:NLSE}, it gives an equation for the potential
\begin{equation}
   V(r)
   =
  \Bigl [\,
     \frac{u''}{u} + \frac{(d-1)}{r} \frac{u'}{u} \,
  \Bigr ]_{\mathrm{sub}}
  +
  g \, A^{2 \kappa} \, u^{2 \kappa}\>,
\end{equation}
where we have subtracted the derivative terms at $r=0$.  It will be useful 
when discussing stability to rewrite the amplitude $A$ of the exact solution 
in terms of the mass $M$. In general the form of $A^2$ is $A^2 = M/[\,C_1(d) \, \Omega(d) \,]$, 
as we will demonstrate below. Then we can rewrite $V(r)$ in the form. 
\begin{equation} \label{V}
   V(r)
   =
  \Bigl [\,
     \frac{u''}{u} + \frac{(d-1)}{r} \frac{u'}{u} \,
  \Bigr ]_{\mathrm{sub}}
  +
  g \, \Bigl [ \frac{M}{C_1 \, \Omega} \Bigr ]^{\kappa} \, u^{2 \kappa} \>.
\end{equation}
As an example, for a Gaussian initial data
\begin{equation}
   \psi(r,t)
   =
   A \, \rme^{ -  \alpha \, r^2/2 - \rmi \, \omega t} \>,
\end{equation}
we find that the potential is now given by
\begin{equation}
   V(r) 
   = 
   g \, A^{2 \kappa } \, \rme^{- \kappa \, \alpha \, r^2} 
   -
   \alpha \, d
   +
   \alpha ^2 \, r^2
   +
   \omega\>.
\end{equation}
Thus, if we choose 
\begin{equation}
   \omega = \alpha \, d\>,
\end{equation}
we find that the Gaussian is an exact solution provided that 
\begin{equation}
   V(r) 
   = 
   g A^{2 \kappa } e^{- \kappa \alpha  r^2} 
   +
   \alpha^2 \, r^2 \>.
\end{equation}
We can rewrite this in terms of the $mass$ of the solution.  We have
\begin{equation}
   \rho(r,t) =  A^2 \, \rme^{-\alpha \, r^2}\>,
\end{equation}
and
\begin{equation}
   M
   = 
   \Omega(d) \int_0^\infty \dd{r} r^{d-1} \rho(r,t) 
   =  
   A^2 \Bigl ( \frac{\pi}{\alpha} \Bigr )^{d/2} \, \Omega(d) \>,
\end{equation}
where $\Omega(d) = 2 \pi^{d/2} / \Gamma(d/2)$, so that 
\begin{equation}
   V(r) 
   = 
   g \, M^\kappa \,
   \Bigl ( \frac{\alpha}{\pi} \Bigr )^{\kappa d/2} \, \rme^{- \kappa \, \alpha \, r^2} 
   +
   \alpha^2 \, r^2 \>.
\end{equation}
This external potential makes the Gaussian an exact solution of the $d$-dimensional 
NLSE with arbitrary nonlinearity exponent $\kappa$. 

\section{Acknowledgments} 

FC, EGC, and JFD would like to thank the Santa Fe Institute and the Center for Nonlinear 
Studies at Los Alamos National Laboratory for their hospitality. AK is grateful to Indian 
National Science Academy (INSA) for awarding him INSA Senior Scientist position at Savitribai 
Phule Pune University, Pune, India. The work at Los Alamos National Laboratory was carried out 
under the auspices of the U.S. Department of Energy and NNSA under Contract No. DEAC52-06NA25396. 

%
%
\section*{References}
%
%
\bibliography{Bibfile.bib}  
%
%
\end{document}